\documentclass[12pt,draftcls, onecolumn,journal]{IEEEtran}
\usepackage{times}
\usepackage[final]{graphicx}
\usepackage[reqno]{amsmath}
\usepackage{amsfonts}
\usepackage{caption2}
\usepackage{times,amsmath,epsfig}
\usepackage{latexsym,amssymb}
\usepackage{cite}

\def\myQED{\mbox{\rule[0pt]{1.5ex}{1.5ex}}}

\newtheorem{thm}{Theorem}

\newtheorem{cor}{Corollary}
\newtheorem{define}{Definition}

\newcommand{\no}{\nonumber}

\begin{document}

\title{The Relay-Eavesdropper Channel: Cooperation for Secrecy}

\author{
Lifeng Lai and Hesham El Gamal\thanks{The authors are with the ECE
Department at the Ohio State University.
Email:\{lail,helgamal\}@ece.osu.edu.}}
\maketitle 



\begin{abstract}

This paper establishes the utility of user cooperation in
facilitating secure wireless communications. In particular, the
four-terminal relay-eavesdropper channel is introduced and an
outer-bound on the optimal rate-equivocation region is derived.
Several cooperation strategies are then devised and the
corresponding achievable rate-equivocation region are
characterized. Of particular interest is the novel
Noise-Forwarding (NF) strategy, where the relay node sends
codewords independent of the source message to confuse the
eavesdropper. This strategy is used to illustrate the deaf helper
phenomenon, where the relay is able to facilitate secure
communications while being totally ignorant of the transmitted
messages. Furthermore, NF is shown to increase the secrecy
capacity in the reversely degraded scenario, where the relay node
fails to offer performance gains in the classical setting. The
gain offered by the proposed cooperation strategies is then proved
theoretically and validated numerically in the additive White
Gaussian Noise (AWGN) channel.
\end{abstract}

\section{Introduction} \label{sect:intro}

Shannon introduced the notion of information theoretic secrecy in
~\cite{Shannon:BSTJ:49}. The model in~\cite{Shannon:BSTJ:49}
assumed that the transmission is noiseless, and used a key $K$ to
protect the confidential message $W$. Taking the transmission
uncertainty into consideration, Wyner introduced the wiretap
channel in \cite{Wyner:BSTJ:75}. In the three-terminal wiretap
channel, a source wishes to transmit confidential messages to a
destination while keeping the messages as secret as possible from
a wiretapper. The wiretapper is assumed to have an unlimited
computation ability and to know the coding/decoding scheme used in
the main (source-destination) channel. Under the assumption that
the source-wiretapper channel is a degraded version of the main
channel, Wyner characterized the trade-off between the throughput
of the main channel and the level of ignorance of the message at
the wiretapper using the rate-equivocation region concept. Loosely
speaking, the equivocation rate measures the residual ambiguity
about the transmitted message at the wiretapper. If the
equivocation rate at the wiretapper is arbitrarily close to the
information rate, the transmission is called perfectly secure.
Csisz$\acute{a}$r and K\"{o}rner extended this work to the
broadcast channel with confidential messages, where the source
sends common information to both the destination and the
wiretapper, and confidential messages are sent only to the
destination~\cite{Csiszar:TIT:78}.

Our work here is motivated by the fact that if the wiretapper
channel is less noisy than the main channel\footnote{The
source-wiretapper channel is said to be less noisy than the
source-receiver channel, if for every $V\to X\to YZ$, $I(V;Z)\geq
I(V;Y)$, where $X$ is the signal transmitted by the source, $Y,Z$
are the received signal of the receiver and the wiretapper
respectively.}, the perfect secrecy capacity of the channel is
zero~\cite{Csiszar:TIT:78}. In this case, it is {\bf infeasible}
to establish a secure link under Wyner's wiretap channel model.
Our main idea is to exploit user cooperation in facilitating the
transmission of confidential messages from the source to the
destination. More specially, we consider a four-terminal
relay-eavesdropper channel, where a source wishes to send messages
to a destination while leveraging the help of a relay node to hide
those messages from the eavesdropper. The eavesdropper in our
model can be viewed as the wireless counterpart of Wyner's
wiretapper. This model generalizes the relay
channel~\cite{Cover:TIT:79} and the wiretap
channel~\cite{Wyner:BSTJ:75}.

The relay channel without security constraints was studied under
various
scenarios~\cite{Meulen:AAP:71,Cover:TIT:79,Sedonaris:TCOM:03,
Kramer:TIT:05,Xie:TIT:05,Lai:TIT:06,Liang:TIT:05,Laneman:TIT:04,Azarian:TIT:05}.
In most of these works, cooperation strategies were constructed to
increase the transmission rate and/or reliability function. In
this paper, we identify a {\bf novel} role of the relay node in
establishing a secure link from the source to the destination.
Towards this end, several cooperation strategies for the
relay-eavesdropper channel are constructed and the corresponding
achieved rate-equivocation regions are characterized. An
outer-bound on the optimal rate-equivocation region is also
derived. The proposed schemes are shown to achieve a positive
perfect secrecy rate in several scenarios where the secrecy
capacity in the absence of the relay node is zero. Quite
interestingly, we establish the deaf-helper phenomenon where the
relay can help while being totally ignorant of the transmitted
message from the source. Furthermore, we show that the relay node
can aid in the transmission of confidential messages in some
settings where classical cooperation fails to offer performance
gains, e.g., the reversely degraded relay channel. Finally, we
observe that the proposed Noise-Forwarding (NF) is intimately
related with the multiple access channel with security
constraints, as evident in the sequel.

At this point, we wish to differentiate our investigation from
earlier relevant works. The relay channel with confidential
messages was studied in~\cite{Oohama:ITW:01,Oohama:TIT:06}, where
the relay node acts both as an eavesdropper and a helper. In the
model of~\cite{Oohama:TIT:06}, the source sends common messages to
the destination using the help of the relay node, but also sends
private messages to the destination while keeping them secret from
the relay. In contrast with~\cite{Oohama:TIT:06}, the relay node
in our work acts as a trusted ``third-party'' whose sole goal is
to facilitate secure communications (imposing an additional
security constraint on the relay node is also considered in
Section~\ref{sec:gau}). The idea of using a ``third-party'' to
facilitate secure communications also appeared
in~\cite{Csiszar:TIT:00}. Contrary to our work, which considers
{\bf noisy} channels,~\cite{Csiszar:TIT:00} focused on the
generation of common random secret keys at two nodes under the
assist of a third-party using a {\bf noiseless} public discussion
channel. The users then use the secret key to establish a secure
link between the source-destination pair. Other recent works on
secure communications investigated the multiple access channel
(MAC) with confidential messages~\cite{Liu:ISIT:06,Liang:TIT:06},
the multiple access channel with a degraded
wiretapper~\cite{Tekin:TIT:06}, and MIMO secure
communications~\cite{Hero:TIT:03}. In summary, it appears that our
relay-eavesdropper model is fundamentally different from the
models considered in all previous works.

Throughout the paper, upper-case letter $X$ denotes a random
variable, lower-case letter $x$ denotes a realization of the
random variable, calligraphic letter $\mathcal{X}$ denotes a
finite alphabet set. Boldface letter $\mathbf{x}$ denotes a
vector, $\{\cdot\}^T$ denotes transpose and $\{\cdot\}^H$ denotes
conjugate transpose. We also let $[x]^+=\max\{0,x\}$.

The rest of the paper is organized as follows. In
section~\ref{sec:model}, we introduce the system model and our
notations. Section~\ref{sec:schemes} describes the proposed
cooperation strategies and characterizes the corresponding
achievable performance. The rate-equivocation outer-bound is also
developed in this section. In Section~\ref{sec:gau}, we discuss
several examples that illustrate interesting aspects of the
relay-eavesdropper channel. Finally, Section~\ref{sec:con} offers
some concluding remarks and briefly outlines possible venues for
future research.

\section{The Relay-Eavesdropper Channel}\label{sec:model}

\begin{figure}[thb]
\centering
\includegraphics[width=0.5 \textwidth]{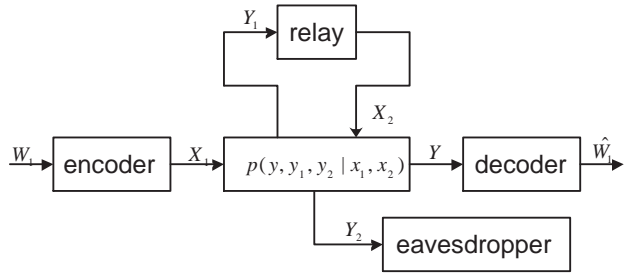}
\caption{The relay eavesdropper channel.} \label{fig:model}
\end{figure}

We consider a four-terminal discrete channel consisting of finite
sets
$\mathcal{X}_{1},\mathcal{X}_{2},\mathcal{Y},\mathcal{Y}_{1},\mathcal{Y}_{2}$
and a transition probability distribution
$p(y,y_{1},y_{2}|x_{1},x_{2})$, as shown in
Figure~\ref{fig:model}. Here, $\mathcal{X}_{1},\mathcal{X}_{2}$
are the channel inputs from the source and the relay respectively,
while $\mathcal{Y},\mathcal{Y}_{1},\mathcal{Y}_{2}$ are the
channel outputs at the destination, relay and eavesdropper
respectively. We impose the memoryless assumption, {\emph i.e.},
the channel outputs $(y_{i},y_{1,i},y_{2,i})$ at time $i$ only
depend on the channel inputs $(x_{1,i},x_{2,i})$ at time $i$. The
source wishes to send the message $W_{1}\in
\mathcal{W}_{1}=\{1,\cdots,M\}$ to the destination using the
$(M,n)$ code consisting: 1) a stochastic encoder $f_n$ at the
source that maps the message $w_{1}$ to a codeword
$\mathbf{x}_{1}\in \mathcal{X}_{1}^{n}$, 2) a relay encoder that
maps the signals $(y_{1,1},y_{1,2},\cdots,y_{1,i-1})$ received
before time $i$ to the channel input $x_{2,i}$, using the mapping
$\varphi_{i}$: $(Y_{1,1}, Y_{1,2}, \cdots, Y_{1,i-1})\rightarrow
X_{2,i}$, 3) a decoding function $\phi$:
$\mathcal{Y}^{n}\rightarrow \mathcal{W}_{1}$. The average error
probability of a $(M,n)$ code is defined as
\begin{eqnarray}
P_{e}^{n}=\sum\limits_{w_{1}\in\mathcal{W}_{1}}\frac{1}{M}\text{Pr}\{\phi(\mathbf{y})\neq
w_{1}|w_{1}\text{ was sent}\}.
\end{eqnarray}
The equivocation rate at the eavesdropper is defined as
\begin{eqnarray}
R_{e}=\frac{1}{n}H(W_{1}|\mathbf{Y}_{2}).
\end{eqnarray}

The rate-equivocation pair $(R_{1},R_{e})$ is said to be
achievable if for any $\epsilon>0$, there exists a sequence of
codes $(M,n)$ such that for any $n\geq n(\epsilon)$, we have
\begin{eqnarray}
R_{1}&=&\frac{1}{n}\log_2M,\\
P_{e}^{n}&\leq& \epsilon,\\
\frac{1}{n}H(W_{1}|\mathbf{Y}_{2})&\geq& R_{e}-\epsilon.
\end{eqnarray}

We further say that the perfect secrecy rate $R_1$ is achievable
if the rate-equivocation pair $(R_1,R_1)$ is achievable. Notice
that if $Y_2=\phi$ (or some other constant), our model reduces to
the classical relay channel without security constraints.

\section{Main Results}\label{sec:schemes}
Our first result establishes an outer-bound on the optimal
rate-equivocation region of the relay-eavesdropper channel.

\begin{thm}\label{thm:up}
In the relay eavesdropper channel, for any rate-equivocation pair
$\{R_1,R_e\}$ with $P_e^n\to 0$ and the equivocation rate at the
eavesdropper larger than $R_e-\epsilon$, there exist some random
variables $U\rightarrow (V_{1},V_{2})\rightarrow
(X_{1},X_{2})\rightarrow (Y,Y_{1},Y_{2})$, such that $(R_1,R_e)$
satisfies the following conditions
\begin{eqnarray}
R_{1}&\leq&\min\{I(V_{1},V_{2};Y),I(V_{1};Y,Y_{1}|V_{2})\},\nonumber\\
R_{e}&\leq& R_{1},\nonumber\\ R_{e}&\leq&
\left[I(V_{1},V_{2};Y|U)-I(V_{1},V_{2};Y_{2}|U)\right]^+.
\end{eqnarray}

\end{thm}
\begin{proof}
Please refer to Appendix~\ref{ap:up}.
\end{proof}
We now turn our attention to constructing cooperation strategies
for the relay-eavesdropper channel. Our first step is to
characterize the achievable rate-equivocation region of Cover-El
Gamal Decode and Forward (DF) Strategy~\cite{Cover:TIT:79}. In DF
cooperation strategy, the relay node will first decode codewords
and then re-encode the message to cooperate with the source. Here,
we use the regular coding and backward decoding scheme developed
in the classical relay setting~\cite{Zing:TIT:89,Kramer:TIT:05},
with the important difference that each message will be associated
with many codewords in order to confuse the eavesdropper.
\begin{thm}\label{thm:df}
The rate pairs in the closure of the convex hull of  all
$(R_{1},R_{e})$ satisfying
\begin{eqnarray}
R_{1}&<&\min \{I(V_{1},V_{2};Y),I(V_{1};Y_{1}|V_{2})\},\nonumber\\
R_{e}&<&R_{1},\\
R_{e}&<&\left[\min
\{I(V_{1},V_{2};Y),I(V_{1};Y_{1}|V_{2})\}-I(V_{1},V_{2};Y_{2})\right]^+,\nonumber
\end{eqnarray}
for some distribution
$p(v_{1},v_{2},x_{1},x_{2},y_{1},y_{2},y)=p(v_1,v_2)p(x_{1},x_{2}|v_1,v_2)p(y_{1},y_{2},y|x_{1},x_{2})$,
are achievable using the DF strategy.

Hence, for the DF scheme, the following perfect secrecy rate is
achievable
\begin{eqnarray}
R_{s}^{(DF)}=\sup\limits_{p(v_1,v_2)}\left[\min
\{I(V_{1},V_{2};Y),I(V_{1};Y_{1}|V_{2})\}-I(V_{1},V_{2};Y_{2})\right]^+.
\end{eqnarray}

\end{thm}
\begin{proof}
Please refer to Appendix~\ref{ap:df}.
\end{proof}


The channel between the source and the relay becomes a {\bf
bottleneck} for the DF strategy when it is noisier than the
source-destination channel. This motivates our Noise-Forwarding
(NF) scheme, where the relay node does not attempt to decode the
message but sends codewords that are {\em independent} of the
source's message. The enabling observation behind this scheme is
that, in the wiretap channel, in addition to its own information,
the source should send extra codewords to confuse the wiretapper.
In our setting, this task can be accomplished by the relay by
allowing it to send independent codewords, which aid in confusing
the eavesdropper.
\begin{figure}[thb]
\centering
\includegraphics[width=0.6 \textwidth]{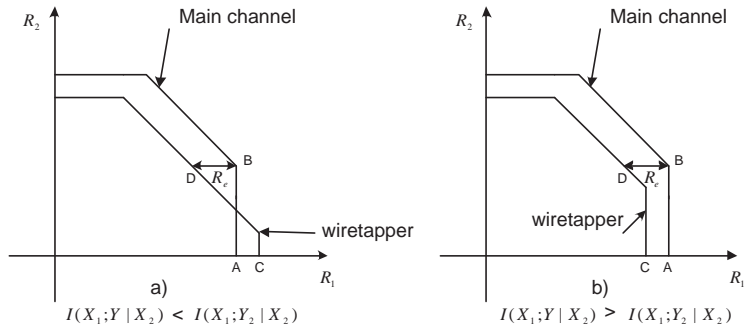}
\caption{The rate region of the compound MACs of the relay
eavesdropper channel for a fixed input distribution
$p(x_1)p(x_2)$.} \label{fig:cmmac}
\end{figure}

Our NF scheme transforms the relay-eavesdropper channel into a
compound multiple access channel (MAC), where the source/relay to
the receiver is the first MAC and source/relay to the eavesdropper
is the second one. Figure~\ref{fig:cmmac} shows the rate region of
these two MACs for a fixed input distribution $p(x_{1})p(x_{2})$.
In the figure, $R_{1}$ is the codeword rate of the source, and
$R_{2}$ is the codeword rate of the relay. We can observe from
Figure 2a) that if the relay node does not transmit, the perfect
secrecy rate is zero for this input distribution since
$R_{1}(A)<R_{1}(C)$. On the other hand, if the relay and the
source coordinate their transmissions and operate at point $B$, we
can achieve the equivocation rate $R_{e}$, which is strictly
larger than zero. On the other hand, in Figure 2b), we can still
get a positive perfect secrecy rate by operating at point $A$ in
the absence of the relay. But by moving the operating point to
$B$, we can get a larger secrecy rate. This illustrates the main
idea of our Noise-Forwarding scheme. The next result establishes
the achievable rate-equivocation region for the NF scheme.
\begin{thm}\label{thm:nf}
The rate pairs in the closure of the convex hull of  all
$(R_{1},R_{e})$ satisfying
\begin{eqnarray}\label{eq:nf}
R_{1}&<& I(V_{1};Y|V_{2}),\nonumber\\
R_{e}&<& R_{1},\\
R_{e}&<&
\left[I(V_{1};Y|V_{2})+\min\{I(V_{2};Y),I(V_{2};Y_{2}|V_{1})\}-\min\{I(V_{2};Y),I(V_2;Y_2)\}-I(V_{1};Y_{2}|V_{2})\right]^+,\nonumber
\end{eqnarray}
for some distribution
$p(v_1,v_2,x_{1},x_{2},y_{1},y_{2},y)=p(v_1)p(v_2)p(x_{1}|v_1)p(x_{2}|v_2)p(y_{1},y_{2},y|x_{1},x_{2})$,
are achievable using the NF scheme.

Hence, for the NF scheme, the achievable perfect secrecy rate is
\begin{eqnarray}
R_{s}^{(NF)}&=&\sup\limits_{p(v_{1})p(v_{2})}\left[I(V_{1};Y|V_{2})+\min\{I(V_{2};Y),I(V_{2};Y_{2}|V_{1})\}\right.\nonumber\\
&&\left.\hspace{15mm}-\min\{I(V_{2};Y),I(V_2;Y_2)\}-I(V_{1};Y_{2}|V_{2})\right]^+.
\end{eqnarray}
\end{thm}
\begin{proof}
Please refer to Appendix~\ref{ap:nf}.
\end{proof}

The following comments are now in order.
\begin{enumerate}
\item The NF scheme is customized to the relay channel with
security constraints which make the transmission of codewords that
are independent of the source message reasonable. Also, in the NF
scheme, the relay node does not need to listen to the source, and
hence, this scheme is more suited works for relay nodes limited by
the half-duplex
constraint~\cite{Laneman:TIT:04,Lai:TIT:06,Khojastepour:TIT:05}.

\item In NF cooperation, each user sends independent messages to
the destination, which resembles the MAC. Hence, NF cooperation
can be adapted to the multiple access eavesdropper channel where
the multiple users in the MAC channel can help each other in
communicating securely with the destination without listening to
each other (note that the results in~\cite{Tekin:TIT:06} were
limited only to the case where the eavesdropped channel is a
degraded version of the channel seen by the destination). Our
related results will be reported elsewhere.

\end{enumerate}


Now, we study another cooperation scheme that does not require
decoding at the relay: Compress and Forward (CF). The CF
cooperation strategy can be viewed as a generalization of NF
where, in addition to the independent codewords, the relay also
sends a quantized version of its noisy observations to the
destination. This noisy version of the relay's observations helps
the destination in decoding the source's message, while the
independent codewords help in confusing the eavesdropper. The
following result establishes the achievable rate-equivocation pair
in the case when $I(X_{1};\hat{Y}_{1},Y|X_{2})\leq
I(X_{1};\hat{Y}_{1},Y_{2}|X_{2})$, {\emph i.e.,} the
source-eavesdropper channel is better than the source-receiver
channel, a situation of particular interest to us.


\begin{thm}\label{thm:cf}
The rate pairs in the closure of the convex hull of  all
$(R_{1},R_{e})$ satisfying
\begin{eqnarray}
R_{1}&<&I(X_{1};\hat{Y}_{1},Y|X_{2}),\no\\
R_{e}&<&R_{1},\\
R_{e}&<&\left[R_{0}+I(X_{1};\hat{Y}_{1},Y|X_{2})-I(X_{1},X_{2};Y_{2})\right]^+,\no
\end{eqnarray}
subject to
\begin{eqnarray}\label{eq:constraint}
\min\{I(X_{2};Y),I(X_2;Y_2|X_1)\}-R_{0}\geq
I(Y_{1};\hat{Y}_{1}|X_{2}),
\end{eqnarray}
for some distribution $p(x_{1},x_{2},y_{1},y_{2},y,\hat{y}_{1})
=p(x_{1})p(x_{2})p(y_{1},y_{2},y|x_{1},x_{2})p(\hat{y}_{1}|y_{1},x_{2}),$
are achievable using CF strategy.
\end{thm}

\begin{proof}
Please refer to Appendix~\ref{ap:cf}.
\end{proof}
Three comments are now in order.

\begin{enumerate}

\item In Theorem~\ref{thm:cf}, $R_{0}$ is the rate of pure noise
generated by the relay to confuse the eavesdropper, while
$\min\{I(X_{2};Y),I(X_{2};Y_{2}|X_{1})\}-R_{0}$ is the part of the
rate allocated to send the compressed signal $\hat{Y}_{1}$ to help
the destination. If we set
$R_{0}=\min\{I(X_{2};Y),I(X_{2};Y_{2}|X_{1})\}$, this scheme
becomes the NF scheme.


\item In order to enable analytical tractability, the
coding/decoding scheme used in the proof is slightly different
from that of~\cite{Cover:TIT:79}. In~\cite{Cover:TIT:79}, the
destination uses sliding-window decoding, while our proof uses
backward decoding. Hence, the bound for $R_e$ provided here is a
lower-bound for the $R_e$ achieved by the CF scheme. One may be
able to achieve a larger $R_e$ using exactly the CF scheme
proposed in~\cite{Cover:TIT:79}. But, unfortunately, we are not
yet able to bound $R_e$ when sliding-window decoding is used.

\item Compared with CF decoding, the proposed NF strategy enjoys
the advantage of simplicity. Also, if one only focuses on the
perfect secrecy rate, it is easy to see that these two schemes
achieve identical performance. Again, this observation is limited
to our lower bound on $R_e$ in Theorem~\ref{thm:cf}.

\end{enumerate}

\section{Examples}\label{sec:gau}
This section discusses several examples that illustrate some
unique features of the relay-eavesdropper channel. For simplicity,
we only focus on the perfect secrecy rate of various schemes.

\subsection{The Deaf Helper Phenomenon}

The security constraints imposed on the network bring about a new
phenomenon which we call the \emph{deaf helper phenomenon}, where
the relay node can still help even it is totally ignorant of the
message transmitted from the source. In this setup, we impose an
additional security constraint on the relay node, and say a rate
$R_{s}$ is achievable for a deaf helper if for any $\epsilon>0$,
there exists a sequence of codes $(M,n)$ such that for any $n\geq
n(\epsilon)$, we have
\begin{eqnarray}
R_{s}&=&\frac{1}{n}\log_2M,\hspace{5mm}
P_{e}^{n}\leq \epsilon,\no\\
\frac{1}{n}H(W_{1}|\mathbf{Y}_{2})&\geq&
R_{s}-\epsilon,\hspace{5mm}
\frac{1}{n}H(W_{1}|\mathbf{Y}_1,\mathbf{X}_2)\geq R_s-\epsilon.
\end{eqnarray}

In this case, the signal received by the relay node does not leak
any information about the transmitted message $W_1$. This model
describes a more conservative scenario where the source does not
trust the relay but still wishes to exploit the benefit brought by
cooperation. We assume that the relay node is not malicious and,
hence, is willing to cooperate with the source\footnote{If the
relay node is malicious, it can then send signals that are
dependent with signal received and then could even block the
transmission of the main channel.}. The following theorem
characterizes the achievable perfect secrecy rate of the NF
strategy in the deaf-helper setting.

\begin{thm}\label{thm:nfdeaf}
The perfect secrecy rate of the NF scheme with an additional
security constraint on the relay node is
$R_s=\max\limits_{p(v_1)p(v_2)}\min\{R_{s1},R_{s2}\},$ where
\begin{eqnarray}R_{s1}&=&\big[I(V_{1};Y|V_{2})+\min\{I(V_{2};Y),I(V_{2};Y_{2}|V_{1})\}-\min\{I(V_{2};Y),I(V_2;Y_2)\}-I(V_{1};Y_{2}|V_{2})\big]^+,
\no\\\no
R_{s2}&=&[I(V_{1};Y|V_{2})-I(V_{1};Y_1|X_{2})]^+.\end{eqnarray}

\end{thm}
\begin{proof}
Please refer to Appendix~\ref{ap:nfdeaf}.
\end{proof}

\subsection{The Reversely Degraded Relay-Eavesdropper Channel}

In the classical relay channel without security constraints, there
exist some scenarios where the relay node does not provide any
gain, for example, the reversely degraded relay channel shown
in~\cite{Cover:TIT:79}. Here, we focus on this scenario and show
that the relay node can still offer a gain in the presence of the
eavesdropper.

\begin{define}[\cite{Cover:TIT:79}]
The relay channel is called reversely degraded, if
$p(y,y_1|x_1,x_2)=p(y|x_1,x_2)$ $p(y_1|y,x_2).$
\end{define}

The following result, borrowed from~\cite{Cover:TIT:79}, states
the capacity of the classical reversely degraded relay channel.
\begin{thm}[Theorem 2,~\cite{Cover:TIT:79}]
The capacity of the reversely degraded relay channel is
\begin{eqnarray}
C_{0}=\max\limits_{x_{2}}\max\limits_{p(x_1)}I(X_1;Y|x_2).
\end{eqnarray}
\end{thm}

This result implies that the relay node should send a constant,
and hence, does not contribute new information to the destination.
In most channel models, the constant sent by the relay does not
result in any capacity gain. The question now is whether the same
conclusion holds in the presence of an eavesdropper.  We first
observe that the degradedness of the relay channel implies that DF
and CF cooperation will not provide the destination with
additional useful information. The relay node, however, can still
send codewords independent of the received signal to confuse the
eavesdropper, as proposed in the NF scheme. Since we do not
require decoding at the relay node in the proof of
Theorem~\ref{thm:nf}, the degradedness imposed here does not
affect the performance. Hence, we get the following achievable
perfect secrecy rate for the reversely degraded relay-eavesdropper
channel.

\begin{cor}
The achievable perfect secrecy rate of the reversely degraded
relay eavesdropper channel is
\begin{eqnarray}
R_{s}&=&\max\limits_{p(v_1)p(v_2)}\big[I(V_{1};Y|V_{2})+\min\{I(V_{2};Y),I(V_{2};Y_{2}|V_{1})\}\no\\
&&\hspace{0.7in}-\min\{I(V_{2};Y),I(V_2;Y_2)\}-I(V_{1};Y_{2}|V_{2})\big]^+.
\end{eqnarray}
\end{cor}

\subsection{The AWGN Channel}
Now we consider the Gaussian relay-eavesdropper channel, where the
signal received at each node is
$$y_{j}[n]=\sum\limits_{i\neq j}h_{ij}x_{i}[n]+z_{j}[n],$$
here $h_{ij}$ is the channel coefficient between node
$i\in\{s,r\}$ and node $j\in\{r,w,d\}$, and $z_{j}$ is the i.i.d
Gaussian noise with unit variance at node $j$. The source and the
relay have average power constraint $P_{1},P_{2}$ respectively.

In~\cite{Leung:TIT:78}, it was shown that the secrecy capacity of
the degraded Gaussian wiretap channel is $[C_M-C_{MW}]^+$, where
$C_M,C_{MW}$ are the capacity of the main channel and wiretap
channel, respectively. This result is also shown to be valid for
stochastically degraded channel~\cite{Liang:TIT:06}. In our case,
if the relay does not transmit, the relay eavesdropper channel
becomes a Gaussian eavesdropper channel, which can always be
converted into a stochastically degraded channel as done in the
Gaussian broadcast channel~\cite{Cover:Book:91}. Applying this
result to our case, the secrecy capacity of the Gaussian
eavesdropper channel without the relay node is given by
$\left[\frac{1}{2}\log_2(1+|h_{sd}|^2P_1)-\frac{1}{2}\log_2(1+|h_{sw}|^2P_1)\right]^+$.
Hence if $|h_{sw}|^2\geq |h_{sd}|^2$ and the relay does not
transmit, the secrecy capacity is zero, no matter how large $P_1$
is. On the other hand, as shown later, the relay can facilitate
the source-destination pair to achieve a positive perfect secrecy
rate under some conditions even when $|h_{sw}|^2\geq |h_{sd}|^2$.
In the following, we focus on such scenarios.

\subsubsection{DF and NF}

At this point, we do not know the optimal input distribution that
maximizes $R_{s}^{(DF)}$, $R_{s}^{(NF)}$. Here, we let $V_1=X_1,
V_2=X_2$ and use a Gaussian input distribution to obtain an
achievable lower bound.

For DF cooperation scheme, we let $X_{2}\sim
\mathcal{N}(0,P_{2})$, $X_{10}\sim \mathcal{N}(0, P)$, where
$\mathcal{N}(0,P)$ is the Gaussian distribution with zero mean and
variance $P$. Also, we let
$$X_{1}=c_{1}X_{2}+X_{10},$$
where $c_{1}$ is a constant to be specified later. In this
relationship, the novel information is modelled by $X_{10}$,
whereas $X_2$ represents the part of the signal which the source
and the relay cooperate in beamforming towards the destination. To
satisfy the average power constraint at the source, we require
$|c_{1}|^2 P_{2}+P\leq P_{1}.$

Straightforward calculations result in
\begin{eqnarray}
I(X_{1};Y_{1}|X_{2})&=&\frac{1}{2}\log_{2}(1+|h_{sr}|^2P),\nonumber\\
I(X_{1},X_{2};Y)&=&\frac{1}{2}\log_{2}(1+|h_{sd}c_{1}+h_{rd}|^2 P_{2}+|h_{sd}|^2P),\nonumber\\
I(X_{1},X_{2};Y_{2})&=&\frac{1}{2}\log_{2}(1+|h_{sw}c_{1}+h_{rw}|^2
P_{2}+|h_{sw}|^2P).\nonumber
\end{eqnarray}
Hence, we have
\begin{eqnarray}\label{eq:dfgau}
R_{s}^{(DF)}=\max
\limits_{c_{1},P}\left[\min\Big\{\frac{1}{2}\log_{2}\Big(\frac{1+|h_{sr}|^2P}{1+|h_{sw}c_{1}+h_{rw}|^2
P_{2}+|h_{sw}|^2P}\Big),\right.\nonumber\\\left.\frac{1}{2}\log_{2}\Big(\frac{1+|h_{sd}c_{1}+h_{rd}|^2
P_{2}+|h_{sd}|^2P}{1+|h_{sw}c_{1}+h_{rw}|^2
P_{2}+|h_{sw}|^2P}\Big)\Big\}\right]^+.
\end{eqnarray}

For NF, we let $X_{1}\sim \mathcal{N}(0, P_{1})$,
$X_{2}\sim\mathcal{N}(0,P_{2})$. Here $X_{1},X_{2}$ are
independent, resulting in
\begin{eqnarray}
&&I(X_{1};Y|X_{2})=\frac{1}{2}\log_{2}\left(1+|h_{sd}|^2
P_{1}\right),\nonumber\\
&&I(X_{1},X_{2};Y)-I(X_{1},X_{2};Y_{2})=\frac{1}{2}\log_{2}\left(\frac{1+
|h_{sd}|^2P_{1}+|h_{rd}|^2 P_{2}}{1+|h_{sw}|^2P_{1}+|h_{rw}|^2 P_{2}}\right),\nonumber\\
&&I(X_{2};Y_{2}|X_{1})+I(X_{1};Y|X_{2})-I(X_{1},X_{2};Y_{2})
=\frac{1}{2}\log_{2}\left(\frac{(1+|h_{rw}|^2
P_{2})(1+|h_{sd}|^2P_{1})}{1+|h_{sw}|^2P_{1}+|h_{rw}|^2
P_{2}}\right).\nonumber
\end{eqnarray}
Hence, we have
\begin{eqnarray}
R_{s}^{(NF)}&=&\left[\min\left\{\frac{1}{2}\log_{2}\left(1+|h_{sd}|^2
P_{1}\right),\frac{1}{2}\log_{2}\left(\frac{1+|h_{sd}|^2P_{1}+|h_{rd}|^2
P_{2}}{1+|h_{sw}|^2P_{1}+|h_{rw}|^2
P_{2}}\right),\right.\right.\\\nonumber&&\left.\left.\hspace{1in}\frac{1}{2}\log_{2}\left(\frac{(1+|h_{rw}|^2
P_{2})(1+|h_{sd}|^2P_{1})}{1+|h_{sw}|^2P_{1}+|h_{rw}|^2
P_{2}}\right)\right\}\right]^+.
\end{eqnarray}

\subsubsection{Amplify and Forward}

In this subsection, we quantify the achievable secrecy rate of
Amplify and Forward (AF) cooperation\footnote{ We did not consider
this scheme in the discrete case since, in general, it does not
lend itself to a single letter characterization.}. In AF, the
source encodes its messages into codewords with length $ML$ each,
and divides each codeword into $L$ sub-blocks each with $M$
symbols, where $L$ is chosen to be sufficiently large. At each
sub-block, the relay sends a linear combination of the received
noisy signal of this sub-block so far. For simplicity, we limit
our discussion to $M=2$. In this case, the source sends $X_{1}(1)$
at the first symbol interval of each sub-block, the relay receives
$Y_{1}(1)=h_{sr}X_{1}(1)+Z_{1}(1)$; At the second symbol interval,
the source sends $\alpha X_{1}(1)+\beta X_{1}(2)$, while the relay
sends $\gamma Y_{1}(1)$. Here $\alpha, \beta, \gamma$ are chosen
to satisfy the average power constraints of the source and the
relay. Thus, this scheme allows beam-forming between the source
and relay without requiring the relay to fully decode.

Writing the signal received at the destination and the
eavesdropper in matrix form, we have
\begin{eqnarray}\label{eq:channel}
\mathbf{Y}=\mathbf{H}_{1}\mathbf{X}_{1}+\mathbf{Z},\quad
\mathbf{Y}_{2}=\mathbf{H}_{2}\mathbf{X}_{1}+\mathbf{Z}_{2},
\end{eqnarray}
where
\begin{eqnarray}
\mathbf{H}_{1}={\small\left[\begin{array}{cc}h_{sd}&0\\\beta
h_{sd}+\gamma h_{sr}h_{rd}&\alpha
h_{sd}\end{array}\right]},\mathbf{H}_{2}={\small\left[\begin{array}{cc}h_{sw}&0\\\beta
h_{sw}+\gamma h_{sr} h_{rw}&\alpha
h_{sw}\end{array}\right]},\nonumber
\end{eqnarray}
\begin{eqnarray}
&&\mathbf{X}_{1}=[X_{1}(1),X_{1}(2)]^T, \mathbf{Z}=[Z(1),\gamma
h_{rd} Z_{1}(1)+Z(2)]^T,\mathbf{Z}_{2}=[Z_{2}(1),\gamma h_{rw}
Z_{1}(1)+Z_{2}(2)]^T,\nonumber\\ &&\mathbf{Y}=[Y(1),Y(2)]^T,
\mathbf{Y}_2=[Y_2(1),Y_{2}(2)]^T.
\end{eqnarray}
The channel under consideration can be viewed as an equivalent
standard memoryless eavesdropper channel with input
$\mathbf{X}_{1}$ and outputs $\mathbf{Y},\mathbf{Y}_{2}$ at the
destination and the eavesdropper respectively. Then, based on the
result of~\cite{Csiszar:TIT:78}, an achievable perfect secrecy
rate is $
[I(\mathbf{X}_{1};\mathbf{Y})-I(\mathbf{X}_{1};\mathbf{Y}_{2})]^+.$

Choosing a Gaussian input with covariance matrix
$\mathbb{E}\{\mathbf{X}\mathbf{X}^H\}=P\mathbf{I},$
 where
$\mathbf{I}$ is the identity matrix, we get the following perfect
secrecy rate
\begin{eqnarray}\label{eq:afgau}
R_{s}^{(AF)}&=&\max\limits_{\alpha,\beta,\gamma,P}
\left[\frac{1}{4}\log_{2}\frac{|\det\{P\mathbf{H}_{1}\mathbf{H}_{1}^H+\mathbb{E}\{\mathbf{Z}\mathbf{Z}^H\}\}|}
{|\det\{\mathbb{E}\{\mathbf{Z}\mathbf{Z}^H\}\}|}
-\frac{1}{4}\log_{2}\frac{|\det\{P\mathbf{H}_{2}\mathbf{H}_{2}^H+\mathbb{E}\{\mathbf{Z}_{2}\mathbf{Z}_{2}^H\}\}|}
{|\det\{\mathbb{E}\{\mathbf{Z}_{2}\mathbf{Z}_{2}^H\}\}|}\right]^+\nonumber\\
&=&\max\limits_{\alpha,\beta,\gamma,P}\left[\frac{1}{4}\log_{2}
\frac{|\det\{P\mathbf{H}_{1}\mathbf{H}_{1}^H+\mathbf{A}\}\det\mathbf{B}|}
{|\det\{P\mathbf{H}_{2}\mathbf{H}_{2}^H+\mathbf{B}\}\det\mathbf{A}|}\right]^+,
\end{eqnarray}
where
\begin{eqnarray}
\mathbf{A}=\small{\left[\begin{array}{cc}1&0\\0&1+|\gamma
h_{rd}|^2\end{array}\right]},\mathbf{B}=\left[\begin{array}{cc}1&0\\0&1+|\gamma
h_{rw}|^2\end{array}\right],\nonumber
\end{eqnarray}
and the maximization is over the set of power constraints:
\begin{eqnarray}
(1+|\alpha|^2+|\beta|^2)P\leq 2P_{1},\nonumber\\
|\gamma|^2(|h_{sr}|^2P+1)\leq 2P_{2}.
\end{eqnarray}
\subsubsection{Numerical Results}

\begin{figure}[thb]
\centering
\includegraphics[width=0.4 \textwidth]{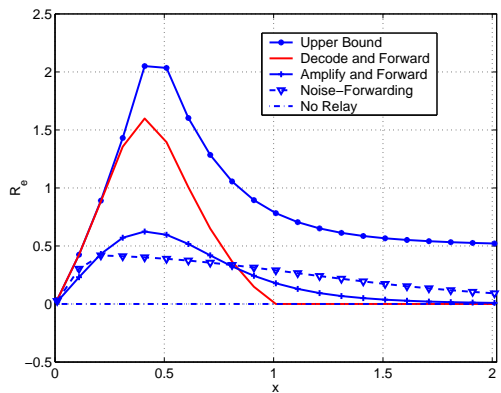}
\caption{The achievable perfect secrecy rate of the proposed
schemes in the Gaussian relay eavesdropper channel.}
\label{fig:gaussianwithcf}
\end{figure}

In this subsection, we give numerical results under two channel
models. The first is the real channel where
$h_{ij}=d_{ij}^{-\gamma}$, with $d_{ij}$ being the distance
between node $i$ and $j$ and $\gamma>1$ is the channel attenuation
coefficient. In the second model, we assume that each channel
experiences an independent phase fading, that is
$h_{ij}=d_{ij}^{-\gamma}e^{j\theta_{ij}}$, where $\theta_{ij}$ is
uniformly distributed over $[0,2\pi)$. We believe that the second
model is more practically relevant than the real channel scenario.

\begin{figure}[thb]
\centering
\includegraphics[width=0.4 \textwidth]{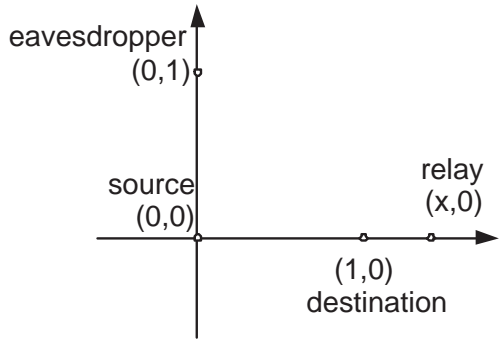}
\caption{The network topology.} \label{fig:networkexam}
\end{figure}

Figure~\ref{fig:gaussianwithcf} shows the achievable perfect
secrecy rate of the proposed schemes for the first channel model.
In generating this figure, we use the network topology shown in
Figure~\ref{fig:networkexam}, where we put the source at $(0,0)$,
the destination at $(1,0)$, the eavesdropper at $(0,1)$, and the
relay node at $(x,0)$. We let $P_{1}=1,P_{2}=8$. Since
$d_{sd}=d_{sw}$, the perfect secrecy capacity of the eavesdropper
channel without the relay node is zero. But, as shown in the
figure, we can achieve a positive secrecy rate by introducing a
relay node. In computing the upper-bound, we set $V_{1}\sim
\mathcal{N}(0,P_{1}),V_{2}\sim \mathcal{N}(0,P_{2})$ with a
correlation coefficient $\rho$, and maximize over $\rho\in[-1,1]$.
Notice that the Gaussian input is not necessarily optimal for the
upper-bound. We can see that, when the relay is near the source,
the DF scheme touches the Gaussian upper-bound. Also, when $x>1$,
it is clear that DF cooperation does not offer any gain, while NF
and AF still offer positive rates. Notice that when $x>1$, both
$d_{sr}$ ,$d_{sd}$ are larger than $d_{sw}$. The interesting
observation here is that though both the destination and relay are
in disadvantage positions compared with the eavesdropper, they can
cooperate with each other and gain some advantage over the
eavesdropper. If the relay is at $0$, our model is equivalent to
the case where the source has two antennas. Notice that the
upper-bound of the perfect secrecy capacity is zero under this
scenario. Hence, increasing the number of transmitting antenna at
the source does not increase the secrecy capacity under the real
channel model. On the other hand, if there is a relay node at an
appropriate position, we can exploit this relay node to establish
a secure source-destination link.
\begin{figure}[thb]
\centering
\includegraphics[width=0.4 \textwidth]{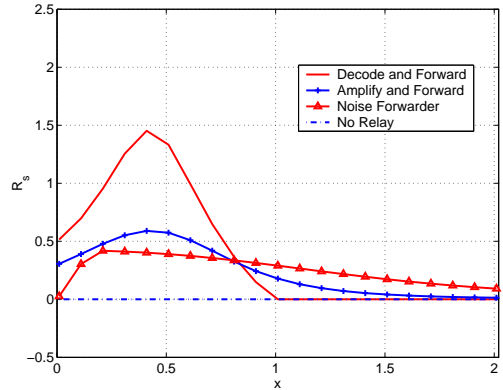}
\caption{The achievable perfect secrecy capacity for various
schemes in the Gaussian relay eavesdropper channel with phase
fading.} \label{fig:gaussianwithcfwp}
\end{figure}

In the second scenario, we assume that before transmission, the
source knows the phases $\theta_{sr},\theta_{sd},\theta_{rd}$, but
does not know $\theta_{sw},\theta_{rw}$. The random phase will not
affect the achievable perfect secrecy rate of NF since it does not
depend on beam-forming between the source and relay. But, the
rates of DF and AF are different here. In both cases, the source
can adjust its phase according to the knowledge of the phase
information about $\theta_{sr},\theta_{sd},\theta_{rd}$. In this
way, the signals of the source and the relay will add up
coherently at the destination, but not at the eavesdropper since
$\theta_{sw},\theta_{rw}$ are independent of
$\theta_{sd},\theta_{rd},\theta_{sr}$. The secrecy rate of DF and
AF could then be obtained by
averaging~\eqref{eq:dfgau},~\eqref{eq:afgau} over the random
phases. Figure~\ref{fig:gaussianwithcfwp} shows the achievable
perfect secrecy rates of the proposed strategies for the same
setup as the first scenario. Due to the random phases, the
achievable perfect secrecy capacity when the relay is at the same
position as the source is not zero anymore. In this case, it will
be beneficial to have multiple transmitting antennas at the
source. Similar to the first scenario, when $x>1$, DF cooperation
does not offer any benefit. But both NF and AF still enjoy
non-zero secrecy rates.
\section{Conclusions}\label{sec:con}
In this paper, the relay-eavesdropper channel was studied. In
particular, several cooperation strategies were proposed and the
corresponding achievable performance bounds were obtained.
Furthermore, an outer-bound on the optimal rate-equivocation
region for this channel was developed. Of particular interest is
the proposed NF strategy which was used to illustrate the
deaf-helper phenomenon, and to demonstrate the utility of the
relay node in the reversely degraded relay-eavesdropper channel.
Overall, our results establish the critical role of user
cooperation in facilitating secure wireless communications and
shed light on the unique feature of the relay-eavesdropper
channel.

Among the many open problems posed by our work, how to close the
gap between the achievable performance and the outer-bound is
arguably the most important one. This problem is expected to be
challenging since the capacity of the classical relay channel
remains unknown. The investigation of the role of feedback in the
relay-eavesdropper channel is another interesting problem. In the
relay channel without security constraints, noiseless/noisy
feedback was shown to be beneficial. On the other hand, in the
presence of an eavesdropper, the role and optimal mechanism of
feedback is not yet known, since the eavesdropper could also
benefit from the feedback signal. Finally, extending our work to a
large scale network is expected to be of practical significance.

\appendices
\section{Proof of Theorem~\ref{thm:up}}\label{ap:up}
The proof follows that of~\cite{Csiszar:TIT:78}.
\begin{eqnarray}
nR_{e}&=& H(W_{1}|Y_{2}^{n})\\
 &=& H(W_{1})-I(W_{1};Y_{2}^{n})\\
 &=&I(W_{1};Y^{n})-I(W_{1};Y_{2}^{n})+H(W_{1}|Y^n)\\
&
\leq&\sum\limits_{i=1}^{n}[I(W_{1};Y_{i}|Y^{i-1})-I(W_{1};Y_{2,i}|Y_{2,i+1}^{n})]+n\delta_{n},
\end{eqnarray}
where $Y^{i-1}=Y(1,\cdots,i-1),Y_{2,i+1}^{n}=Y_{2}(i+1,\cdots,n)$,
and $\delta_{n}\rightarrow 0$ as $n\to \infty$. We get this by
using the chain rule to expand $I(W_{1};Y^{n})$ from $i=1$ and
expand $I(W_{1};Y_{2}^{n})$ from $i=n$, also we use the Fano's
inequality to bound $H(W_{1}|Y^n)$.

We continue
\begin{eqnarray}
nR_{e}&
\leq&\sum\limits_{i=1}^{n}[I(W_{1};Y_{i}|Y^{i-1})-I(W_{1};Y_{2,i}|Y_{2,i+1}^{n})]+n\delta_{n}\\
&=&\sum\limits_{i=1}^{n}[I(W_{1},Y_{2,i+1}^n;Y_{i}|Y^{i-1})-I(Y_{2,i+1}^n;Y_{i}|Y^{i-1},W_1)\\
&&-I(W_{1},Y^{i-1};Y_{2,i}|Y_{2,i+1}^{n})+I(Y^{i-1};Y_{2,i}|Y_{2,i+1}^{n},W_1)]+n\delta_{n}\\
&=&\sum\limits_{i=1}^{n}[I(W_{1},Y_{2,i+1}^n;Y_{i}|Y^{i-1})-I(W_{1},Y^{i-1};Y_{2,i}|Y_{2,i+1}^{n})]+n\delta_{n},
\end{eqnarray}
since 
$\sum\limits_{i=1}^nI(Y_{2,i+1}^n;Y_{i}|Y^{i-1},W_1)=\sum\limits_{i=1}^nI(Y^{i-1};Y_{2,i}|Y_{2,i+1}^{n},W_1),$
which is proved in the lemma 7 of ~\cite{Csiszar:TIT:78}. Now
\begin{eqnarray}
nR_{e}&
\leq&\sum\limits_{i=1}^{n}[I(W_{1},Y_{2,i+1}^n;Y_{i}|Y^{i-1})-I(W_{1},Y^{i-1};Y_{2,i}|Y_{2,i+1}^{n})]+n\delta_{n}\\
&=&\sum\limits_{i=1}^{n}[I(Y_{2,i+1}^n;Y_{i}|Y^{i-1})+I(W_{1};Y_{i}|Y^{i-1},Y_{2,i+1}^n)\nonumber\\
&&-I(Y^{i-1};Y_{2,i}|Y_{2,i+1}^{n})-I(W_{1};Y_{2,i}|Y^{i-1},Y_{2,i+1}^{n})]+n\delta_{n}\\
&=&\sum\limits_{i=1}^{n}[I(W_{1};Y_{i}|Y^{i-1},Y_{2,i+1}^n)-I(W_{1};Y_{2,i}|Y^{i-1},Y_{2,i+1}^{n})]+n\delta_{n},
\end{eqnarray}
since
$\sum\limits_{i=1}^nI(Y_{2,i+1}^n;Y_{i}|Y^{i-1})=\sum\limits_{i=1}^nI(Y^{i-1};Y_{2,i}|Y_{2,i+1}^{n})$,
which is also proved in~\cite{Csiszar:TIT:78}.

Now, let $J$ be a random variable uniformly distributed over
$\{1,\cdots,n\}$, set
$U=JY^{i-1}Y_{2,i+1}^{n},$$V_{1}=JY_{2,i+1}^nW_{1},V_{2}=JY^{i-1},Y_{1}=Y_{1,J},Y_{2}=Y_{2,J}$,
$Y=Y_{J}$, $X_{1}=X_{1,J},X_{2}=X_{2,J}$ we have
\begin{eqnarray} R_{e}&\leq&
\frac{1}{n}\sum\limits_{i=1}^{n}[I(W_{1};Y_{i}|Y^{i-1},Y_{2,i+1}^n)-I(W_{1};Y_{2,i}|Y^{i-1},Y_{2,i+1}^{n})]+\delta_{n}\nonumber\\
&=&
\frac{1}{n}\sum\limits_{i=1}^{n}[I(W_{1},Y^{i-1},Y_{2,i+1}^n;Y_{i}|Y^{i-1},Y_{2,i+1}^n)-I(W_{1},Y^{i-1},Y_{2,i+1}^n;Y_{2,i}|Y^{i-1},Y_{2,i+1}^{n})]+\delta_{n}\nonumber\\
&=&I(V_{1},V_{2};Y|U)-I(V_{1},V_{2};Y_{2}|U)+\delta_{n}.
\end{eqnarray}
Since the channel is memoryless, one can then check that
$U\to(V_1,V_2)\to(X_1,X_2)\to(Y,Y_1,Y_2)$ is a Markov chain. In
the following, we bound $R_{1}$.
\begin{eqnarray}
I(W_{1};\mathbf{Y})&=&\sum\limits_{i=1}^{n}I(W_{1};Y_{i}|Y^{i-1})\nonumber\\
&=&\sum\limits_{i=1}^{n}[H(Y_{i}|Y^{i-1})-H(Y_{i}|W_{1},Y^{i-1})]\nonumber\\
&\leq&\sum\limits_{i=1}^{n}[H(Y_{i})-H(Y_{i}|W_{1},Y^{i-1})]\nonumber\\
&\leq&\sum\limits_{i=1}^{n}[H(Y_{i})-H(Y_{i}|W_{1},Y^{i-1},Y_{2,i+1}^{n})]\nonumber\\
&=&\sum\limits_{i=1}^{n}I(W_{1},Y^{i-1},Y_{2,i+1}^{n};Y_{i}).
\end{eqnarray}
Hence
\begin{eqnarray}
R_{1}\leq \frac{1}{n}I(W_{1};\mathbf{Y})\leq I(V_{1},V_{2};Y).
\end{eqnarray}
Also
\begin{eqnarray}
I(W_{1};\mathbf{Y})&=&\sum\limits_{i=1}^n I(W_{1};Y_{i}|Y^{i-1})\nonumber\\
&\leq&\sum\limits_{i=1}^n
I(W_{1};Y_{i},Y_{1,i}|Y^{i-1})\nonumber\\
&=&\sum\limits_{i=1}^n
[H(Y_{i},Y_{1,i}|Y^{i-1})-H(Y_{i},Y_{1,i}|W_{1},Y^{i-1})]\nonumber\\
&\leq&\sum\limits_{i=1}^n[
H(Y_{i},Y_{1,i}|Y^{i-1})-H(Y_{i},Y_{1,i}|W_{1},Y^{i-1},Y_{2,i+1}^{n})]\nonumber\\
&=&\sum\limits_{i=1}^nI(W_{1},Y_{2,i+1}^n;Y_{i},Y_{1,i}|Y^{i-1}).
\end{eqnarray}
Hence, we have
\begin{eqnarray}
R_{1}\leq \frac{1}{n}I(W_{1};\mathbf{Y})= I(V_{1};Y,Y_{1}|V_{2}).
\end{eqnarray}
So, we have
\begin{eqnarray}
R_{1}\leq \min\{I(V_{1},V_{2};Y),I(V_{1};Y,Y_{1}|V_{2})\}.
\end{eqnarray}
 The claim is proved.
\section{Proof of Theorem~\ref{thm:df}}\label{ap:df}
The proof is a combination of the coding schemes of
Csisz$\acute{a}$r \emph{et. al.}~\cite{Csiszar:TIT:78} and the
regular coding and backward decoding scheme in the relay
channel~\cite{Kramer:TIT:05, Zing:TIT:89}. We first replace
$V_1,V_2$ in Theorem~\ref{thm:df} with $X_1,X_2$. After proving
Theorem~\ref{thm:df} with $V_1,V_2$ replaced by $X_1,X_2$, we then
prefix a memoryless channel with input $V_1,V_2$ and transmission
probability $p(x_1,x_2|v_1,v_2)$ as reasoned
in~\cite{Csiszar:TIT:78} to finish our proof.

\begin{enumerate}
\item \textbf{Codebook generation:}

For a given distribution $p(x_{1},x_{2})$, we first generate at
random $2^{nR}$ i.i.d $n$-sequence at the relay node each drawn
according to $p(\mathbf{x}_{2})=\prod_{i=1}^{n}p(x_{2,i})$, index
them as $\mathbf{x}_{2}(a),a\in[1,2^{nR}]$, where
$R=\min\{I(X_1,X_2;Y),I(X_1;Y_1|X_2)\}-\epsilon_0$. For each
$\mathbf{x}_{2}(a)$, generate $2^{nR}$ conditionally independent
$n$-sequence $\mathbf{x}_{1}(k,a),k\in[1,2^{nR}]$ drawn randomly
according to
$p(\mathbf{x}_{1}|\mathbf{x}_{2}(a))=\prod_{i=1}^{n}p(x_{1,i}|x_{2,i}(a))$.
Define $\mathcal{W}=\{1,\cdots,2^{n[R-I(X_{1},X_{2};Y_{2})]}\}$,
$\mathcal{L}=\{1,\cdots,2^{nI(X_{1},X_{2};Y_{2})}\}$ and
$\mathcal{K}=\mathcal{W}\times\mathcal{L}=\{1,\cdots,2^{nR}\}$.

\item \textbf{Encoding}

We exploit the block Markov coding scheme, as argued
in~\cite{Cover:TIT:79}, the loss induced by this scheme is
negligible as the number of blocks $B\rightarrow\infty$.

For a given rate pair $(R_{1},R_{e})$ with $R_1\leq R$ and $
R_e\leq R_1$, we give the following coding strategy. Let the
message to be transmitted at block $i$ be
$w_{1}(i)\in\mathcal{W}_{1}=\{1,\cdots,M\}$, where $M=2^{nR_{1}}$.

The stochastic encoder at the transmitter first forms the
following mappings.
\begin{itemize}

\item If $R_{1}>R-I(X_{1},X_{2};Y_{2})$, then we let
$\mathcal{W}_{1}=\mathcal{W}\times\mathcal{J}$, where
$\mathcal{J}=\{1,\cdots,2^{n(R_{1}-[R-I(X_{1},X_{2};Y_{2})])}\}$.
We let $g_{1}$ be the partition that partitions $\mathcal{L}$ into
$|\mathcal{J}|$ equal size subsets. The stochastic encoder at
transmitter will choose a mapping for each message
$w_{1}(i)=(w(i),j(i))\rightarrow(w(i),l(i))$, where $l(i)$ is
chosen randomly from the set $g_{1}^{-1}(j(i))\subset\mathcal{L}$
with uniform distribution.

\item If $R_{1}<R-I(X_{1},X_{2};Y_{2})$, the stochastic encoder
will choose a mapping $w_{1}(i)\rightarrow (w_{1}(i),l(i))$, where
$l(i)$ is chosen uniformly from the set $\mathcal{L}$.
\end{itemize}

Assume that the message $w_{1}(i-1)$ transmitted at block $i-1$ is
associated with $(w(i-1),l(i-1))$
 and the message $w_{1}(i)$ intended to send at block $i$ is associated with $(w(i),l(i))$
 by the stochastic encoder at the transmitter. We let $a(i-1)=(w(i-1),l(i-1))$ and $b(i)=(w(i),l(i))$. The
encoder then sends $\mathbf{x}_{1}(b(i),a(i-1))$. The relay has an
estimation $\hat{\hat{a}}(i-1)$ (see the decoding part), and thus
sends the corresponding codeword
$\mathbf{x}_{2}(\hat{\hat{a}}(i-1))$.

At block $1$, the source sends $\mathbf{x}_{1}(b(1),1)$, the relay
sends $\mathbf{x}_{2}(1)$.

At block $B$, the source sends $\mathbf{x}_{1}(1,a(B-1))$, and the
relay sends $\mathbf{x}_{2}(\hat{\hat{a}}(B-1))$.

\item \textbf{Decoding}

At the end of block $i$, the relay already has an estimation of
the $\hat{\hat{a}}(i-1)$, which was sent at block $i-1$, and will
declare that it receives $\hat{\hat{a}}(i)$, if this is the only
pair such that
$(\mathbf{x}_{1}(\hat{\hat{a}}(i),\hat{\hat{a}}(i-1)),\mathbf{x}_{2}(\hat{\hat{a}}(i-1)),\mathbf{y}_{1}(i))$
are jointly typical. Since $R=\min\{I(X_{1};Y_1|X_2),$
$I(X_1,X_2;Y)\}-\epsilon\leq I(X_{1};Y_{1}|X_{2})-\epsilon$, then
based on the AEP, one has $\hat{\hat{a}}(i)=a(i)$ with probability
goes to 1.

The destination decodes from the last block, {\emph i.e.} block
$B$. Suppose that at the end of block $B-1$, the relay decodes
successfully, then the destination will declare that
$\hat{a}(B-1)$ is received, if
$(\mathbf{x}_{1}(1,\hat{a}(B-1)),\mathbf{x}_{2}(\hat{a}(B-1)),\mathbf{y})$
are jointly typical. It's easy to see that if $R\leq
I(X_{1},X_{2};Y)$, we will have $\hat{a}(B-1)=a(B-1)$ with
probability goes to 1, as $n$ increases.

After getting $\hat{a}(B-1)$, the receiver can get an estimation
of $a(i), i\in[1,B-2]$ in a similar way.

Having $\hat{a}(i-1)$, the destination can get the estimation of
the message $w_{1}(i-1)$ by letting

1)
$\hat{w}_{1}(i-1)=(\hat{w}(i-1),\hat{j}(i-1))=(\hat{w}(i-1),g_{1}(\hat{l}(i-1)))$
if $R_{1}>R-I(X_{1},X_{2};Y_{2})$,

2) $\hat{w}_{1}(i-1)=\hat{w}(i-1)$ if
$R_{1}<R-I(X_{1},X_{2};Y_{2})$.

The probability that $\hat{w}_1(i-1)=w_1(i-1)$ goes to one for
sufficiently large $n$.

 \item
\textbf{Equivocation Computation}
\begin{eqnarray}
H(W_{1}|\mathbf{Y}_{2})&=&H(W_{1},\mathbf{Y}_{2})-H(\mathbf{Y}_{2})\nonumber\\
&=&H(W_{1},\mathbf{Y}_{2},\mathbf{X}_{1},\mathbf{X}_{2})-H(\mathbf{X}_{1},\mathbf{X}_{2}|W_{1},\mathbf{Y}_{2})-H(\mathbf{Y}_{2})\nonumber\\
&=&H(\mathbf{X}_{1},\mathbf{X}_{2})+H(W_{1},\mathbf{Y}_{2}|\mathbf{X}_{1},\mathbf{X}_{2})-H(\mathbf{X}_{1},\mathbf{X}_{2}|W_{1},\mathbf{Y}_{2})-H(\mathbf{Y}_{2})\nonumber\\
&\geq&H(\mathbf{X}_{1})+H(\mathbf{Y}_{2}|\mathbf{X}_{1},\mathbf{X}_{2})-H(\mathbf{X}_{1},\mathbf{X}_{2}|W_{1},\mathbf{Y}_{2})-H(\mathbf{Y}_{2}).
\end{eqnarray}

First, let us calculate
$H(\mathbf{X}_{1},\mathbf{X}_{2}|W_{1},\mathbf{Y}_{2})$. Given
$W_{1}$, the eavesdropper can also do backward decoding as the
receiver. At the end of block $B$, given $W_{1}$, the eavesdropper
knows $w(B-1)$, hence it will decode $l(B-1)$, by letting
$l(B-1)=\hat{l}(B-1)$, if $\hat{l}(B-1)$ is the only one such that
$(\mathbf{x}_{1}(1,(w(B-1),\hat{l}(B-1))),\mathbf{x}_{2}((w(B-1),\hat{l}(B-1))),\mathbf{y})$
are jointly typical. Since $l\in[1,2^{nI(X_{1},X_{2};Y_{2})}]$, we
have
\begin{eqnarray}&&\text{Pr}\{(\mathbf{X}_{1}(1,a(i-1)),\mathbf{X}_{2}(\hat{\hat{a}}(i-1)))\nonumber\\&& \hspace{.5in}\neq
(\mathbf{X}_{1}(1,(w(B-1),\hat{l}(B-1))),\mathbf{X}_{2}((w(B-1),\hat{l}(B-1))))\}\leq
\epsilon_{1}.\end{eqnarray}

Then based on Fano's inequality, we have
\begin{eqnarray}
\frac{1}{n}H(\mathbf{X}_{1},\mathbf{X}_{2}|W_{1}=w_{1},\mathbf{Y}_{2})\leq
\frac{1}{n}+\epsilon_{1} I(X_{1},X_{2};Y_{2})
\end{eqnarray}
Hence, we have
\begin{eqnarray}
\frac{1}{n}H(\mathbf{X}_{1},\mathbf{X}_{2}|W_{1},\mathbf{Y}_{2})=\frac{1}{n}\sum\limits_{w_{1}\in\mathcal{W}_{1}}p(W_{1}=w_{1})H(\mathbf{X}_{1},\mathbf{X}_{2}|W_{1}=w_{1},\mathbf{Y}_{2})\leq
\epsilon_{2},
\end{eqnarray}
 when $n$ is sufficiently large.

Since the channel is memoryless, we have
$H(\mathbf{Y}_{2})-H(\mathbf{Y}_{2}|\mathbf{X}_{1},\mathbf{X}_{2})\leq
nI(X_{1},X_{2};Y_{2})+n\delta_n$, where $\delta_n\to 0,$ as $n\to
\infty$~\cite{Wyner:BSTJ:75}.

Now, from the code construction, we have $H(\mathbf{X}_{1})=nR$ if
$R_1>R-I(X_1,X_2;Y_2)$. In this case, we get
$nR_{e}=H(W_{1}|\mathbf{Y}_{2})\geq
n(R-I(X_{1},X_{2};Y_{2})-\epsilon_{3})$. If $R_1\leq
R-I(X_1,X_2;Y_2)$, $H(\mathbf{X}_{1})=n(R_1+I(X_1,X_2;Y_2))$, in
this case, we get the perfect secrecy, since
$$nR_{e}\geq n( R_1+I(X_1,X_2;Y_2))-nI(X_{1},X_{2};Y_{2})-n\epsilon_{3}\geq
n(R_1-\epsilon_3).$$
\end{enumerate}

The claim is proved.
\section{Proof of Theorem~\ref{thm:nf}}\label{ap:nf}

As~\cite{Csiszar:TIT:78}, we first prove the result for the case
where $V_1,V_2$ in Theorem~\ref{thm:nf} are replaced with
$X_1,X_2$, then prefix a memoryless channel with transition
probability $p(x_1|v_1)p(x_2|v_2)$ to finish our proof.

We first consider the case
$I(X_{1};Y|X_{2})<I(X_{1};Y_{2}|X_{2})$, \emph{i.e.}, the channel
between the source and the eavesdropper is better than the channel
between the source and the destination. In this case, we only need
to consider $\min\{I(X_2;Y), I(X_2;Y_2)\}=I(X_2;Y_2)$, otherwise,
the secrecy rate will be zero. Thus in this case, the last
equation in~\eqref{eq:nf} changes to $R_e<
\big[I(X_{1};Y|X_{2})+\min\{I(X_{2};Y),I(X_{2};Y_{2}|X_{1})\}-I(X_{1},X_{2};Y_{2})\big]^+$.
\begin{enumerate}
\item \textbf{Codebook generation:}

For a given distribution $p(x_{1})p(x_{2})$, we generate at random
$2^{nR_{2}}$ i.i.d $n$-sequence at the relay node each drawn
according to $p(\mathbf{x}_{2})=\prod_{i=1}^{n}p(x_{2,i})$, index
them as $\mathbf{x}_{2}(a),a\in[1,2^{nR_{2}}]$, where we set
$R_2=\min\{I(X_2;Y),I(X_2;Y_2|X_1)\}-\epsilon$. We also generate
random $2^{nR}$ i.i.d $n$-sequence at the source each drawn
according to $p(\mathbf{x}_{1})=\prod_{i=1}^{n}p(x_{1,i})$, index
them as $\mathbf{x}_{1}(k),k\in[1,2^{nR}]$, where
$R=I(X_1;Y|X_2)-\epsilon$. Let
$$R^{'}=\min\{I(X_{2};Y),I(X_{2};Y_{2}|X_{1})\}+I(X_{1};Y|X_{2})-I(X_{1},X_{2};Y_{2}),$$
and define $\mathcal{W} = \{1,\cdots,2^{nR^{'}}\},$
$\mathcal{L}=\{1,\cdots,2^{n(R-R^{'})}\}$ and
$\mathcal{K}=\mathcal{W}\times\mathcal{L}=\{1,\cdots,2^{nR}\}$.

\item \textbf{Encoding}

For a given rate pair $(R_{1},R_{e})$ with $R_1\leq R, R_e\leq
R_1$, we give the following coding strategy. Let the message to be
transmitted at block $i$ be $w_{1}(i)\in\mathcal{W}_{1}=[1,M]$,
where $M=2^{nR_{1}}$.

The stochastic encoder at the transmitter first forms the
following mappings.
\begin{itemize}

\item If $R_{1}>R^{'}$, then we let
$\mathcal{W}_{1}=\mathcal{W}\times\mathcal{J}$, where
$\mathcal{J}=\{1,2^{n(R_{1}-R^{'})}\}$. We let $g_{1}$ be the
partition that partitions $\mathcal{L}$ into $|\mathcal{J}|$ equal
size subsets. The stochastic encoder at transmitter will choose a
mapping for each message
$w_{1}(i)=(w(i),j(i))\rightarrow(w(i),l(i))$, where $l(i)$ is
chosen randomly from the set $g_{1}^{-1}(j(i))\subset\mathcal{L}$
with uniform distribution.

\item If $R_{1}<R^{'}$, the stochastic encoder will choose a
mapping $w_{1}(i)\rightarrow (w_{1}(i),l(i))$, where $l(i)$ is
chosen uniformly from the set $\mathcal{L}$.
\end{itemize}

Suppose the message $w_{1}(i)$ intended to send at block $i$ is
associated with $(w(i),l(i))$
 by the stochastic encoder at the transmitter. The
encoder then sends $\mathbf{x}_{1}((w(i),l(i)))$. The relay
uniformly picks a code $\mathbf{x}_{2}(a)$ from
$a\in[1,\cdots,2^{nR_{2}}]$, and sends $\mathbf{x}_{2}(a)$.

\item \textbf{Decoding}

At the end of block $i$, the destination declares that
$\hat{a}(i)$ is received, if $\hat{a}(i)$ is the only one such
that $(\mathbf{x}_{2}(\hat{a}(i)),\mathbf{y})$ are jointly
typical. If there does not exist or there exist more than one such
sequences, the destination declares an error. Since
$R_{2}=\min\{I(X_2;Y),I(X_2;Y_2|X_1)\}-\epsilon\leq
I(X_2;Y)-\epsilon$, then based on AEP, we know that the error
probability will be less than any given positive number
$\epsilon$, when the codeword length $n$ is long enough.

The destination then declares that $\hat{k}$ is received, if
$\hat{k}$ is the only one such that
$(\mathbf{x}_{1}(\hat{k}),\mathbf{x}_{2}(\hat{a}),\mathbf{y})$ are
jointly typical, otherwise declares an error. Since
$R=I(X_1;Y|X_2)-\epsilon$, then based on AEP, we know that we will
have error probability goes to zero, when $n$ is sufficiently
large.

Having $\hat{k}(i)$, the destination can get the estimation of the
message $w_{1}(i)$ by letting

1)
$\hat{w}_{1}(i)=(\hat{w}(i),\hat{j}(i))=(\hat{w}(i),g_{1}(\hat{l}(i)))$,
if $R_{1}>R^{'}$,

2) $\hat{w}_{1}(i)=\hat{w}(i)$, if $R_{1}<R^{'}$.

The probability that $\hat{w}_1(i)=w_1(i)$ goes to one for
sufficiently large $n$.

 \item
\textbf{Equivocation Computation}
\begin{eqnarray}
H(W_{1}|\mathbf{Y}_{2})&=&H(W_{1},\mathbf{Y}_{2})-H(\mathbf{Y}_{2})\nonumber\\
&=&H(W_{1},\mathbf{Y}_{2},\mathbf{X}_{1},\mathbf{X}_{2})-H(\mathbf{X}_{1},\mathbf{X}_{2}|W_{1},\mathbf{Y}_{2})-H(\mathbf{Y}_{2})\nonumber\\
&=&H(\mathbf{X}_{1},\mathbf{X}_{2})+H(W_{1},\mathbf{Y}_{2}|\mathbf{X}_{1},\mathbf{X}_{2})-H(\mathbf{X}_{1},\mathbf{X}_{2}|W_{1},\mathbf{Y}_{2})-H(\mathbf{Y}_{2})\nonumber\\
&\geq&H(\mathbf{X}_{1},\mathbf{X}_{2})+H(\mathbf{Y}_{2}|\mathbf{X}_{1},\mathbf{X}_{2})-H(\mathbf{X}_{1},\mathbf{X}_{2}|W_{1},\mathbf{Y}_{2})-H(\mathbf{Y}_{2}).
\end{eqnarray}

Now let's calculate
$H(\mathbf{X}_{1},\mathbf{X}_{2}|W_{1},\mathbf{Y}_{2})$. Given
$W_{1}$, the eavesdropper can do joint decoding. At any block $i$,
given $W_{1}$, the eavesdropper knows $w(i)$, hence it will decode
$l(i)$ and $a(i)$ sent by the relay, by letting
$l(i)=\hat{l}(i),a(i)=\hat{a}(i)$, if $\hat{l}(i),\hat{a}(i)$ are
the only one such that
$(\mathbf{x}_{1}(w(i),\hat{l}(i)),\mathbf{x}_{2}(\hat{a}(i)),\mathbf{y})$
are jointly typical. Then, since
$R_{2}=\min\{I(X_2;Y),I(X_2;Y_2|X_1)\}-\epsilon\leq
I(X_2;Y_2|X_1)-\epsilon$, we get
\begin{eqnarray}
\frac{1}{2}\log_{2}(|\mathcal{L}|)+R_{2}&=&R+I(X_{1},X_{2};Y_{2})-\min\{I(X_{2};Y),I(X_{2};Y_{2}|X_{1})\}\nonumber\\&&-I(X_{1};Y|X_{2})+\min\{I(X_{2};Y),I(X_{2};Y_{2}|X_{1})\}-\epsilon\nonumber\\&\leq&I(X_{1},X_{2};Y_{2})-\epsilon,
\end{eqnarray}
Also, we have $\frac{1}{2}\log_{2}(|\mathcal{L}|)<R\leq
I(X_{1};Y_{2}|X_{2})-\epsilon.$

So
$\text{Pr}\{(\mathbf{X}_{1}(w(i),\hat{l}(i)),\mathbf{X}_{2}(\hat{a}(i)))\neq
(\mathbf{X}_{1}(w(i),l(i)),\mathbf{X}_{2}(a(i)))\}\leq
\epsilon_{1}.$

Then based on Fano's inequality, we have
\begin{eqnarray}
\frac{1}{n}H(\mathbf{X}_{1},\mathbf{X}_{2}|W_{1}=w_{1},\mathbf{Y}_{2})\leq
\frac{1}{n}+\epsilon_{1} I(X_{1},X_{2};Y_{2})
\end{eqnarray}
Hence, we have
\begin{eqnarray}
\frac{1}{n}H(\mathbf{X}_{1},\mathbf{X}_{2}|W_{1},\mathbf{Y}_{2})=\frac{1}{n}\sum\limits_{w_{1}\in\mathcal{W}_{1}}p(W_{1}=w_{1})H(\mathbf{X}_{1},\mathbf{X}_{2}|W_{1}=w_{1},\mathbf{Y}_{2})\leq
\epsilon_{2},
\end{eqnarray}
 when $n$ is sufficiently large.

Now,
$H(\mathbf{Y}_{2})-H(\mathbf{Y}_{2}|\mathbf{X}_{1},\mathbf{X}_{2})\leq
nI(X_{1},X_{2};Y_{2})+n\delta_n,$ where $\delta_n\to 0,$ as $n\to
\infty$.

Also we have
$H(\mathbf{X}_{1},\mathbf{X}_{2})=H(\mathbf{X}_{1})+H(\mathbf{X}_{2})$
since $\mathbf{x}_{1}$ and $\mathbf{x}_{2}$ are independent. If
$R_{1}>R^{'}$, we have
$H(\mathbf{X}_{1},\mathbf{X}_{2})=R+R_{2}=I(X_{1};Y|X_{2})+\min\{I(X_{2};Y),I(X_{2};Y_{2}|X_{1})\}$.
Combining these, we get $nR_{e}=H(W_{1}|\mathbf{Y}_{2})\geq
n(\min\{I(X_{2};Y),I(X_{2};Y_{2}|X_{1})\}+I(X_{1};Y|X_{2}))-nI(X_{1},X_{2};Y_{2})-n\epsilon_{4}$.

On the other hand, if $R_{1}<R^{'}$, we have
$H(\mathbf{X}_1)=R_1+I(X_1,X_2;Y_2)-\min\{I(X_2;Y),I(X_2;Y_2|X_1)\}$,
hence we have
$H(\mathbf{X}_{1},\mathbf{X}_{2})=R_1+I(X_1,X_2;Y_2)-\epsilon$. We
get perfect secrecy rate, since
$nR_{e}=H(W_{1}|\mathbf{Y}_{2})\geq nR_1-n\epsilon_4$.
\end{enumerate}

This case is proved.

Now, consider the case $I(X_{1};Y|X_{2})>I(X_{1};Y_{2}|X_2)$. If
$\min\{I(X_{2};Y),I(X_2;Y_2)\}=I(X_{2};Y)$, then we have
$\min\{I(X_{2};Y),I(X_2;Y_2|X_1)\}=I(X_{2};Y)$, because
$I(X_2;Y_2|X_1)>I(X_2;Y_2)$ since $X_1,X_2$ are independent. Under
this case, we only need to prove $R_{e}\leq
I(X_1;Y|X_2)-I(X_1;Y_2|X_2)$ are achievable, which can be achieved
by letting the codeword rate be $I(X_1;Y|X_2)$ and
$R^{'}=I(X_1;Y|X_2)-I(X_1;Y|X_2)$. Now the equivocation rate of
the eavesdropper can be calculated as
\begin{eqnarray}\label{eq:ocase}
H(W_{1}|\mathbf{Y}_{2})&\geq&H(W_{1}|\mathbf{Y}_2,\mathbf{X}_2)\no\\&=&H(W_{1},\mathbf{Y}_{2}|\mathbf{X}_2)-H(\mathbf{Y}_{2}|\mathbf{X}_2)\nonumber\\
&=&H(W_{1},\mathbf{Y}_{2},\mathbf{X}_{1}|\mathbf{X}_{2})-H(\mathbf{X}_{1}|W_{1},\mathbf{Y}_{2},\mathbf{X}_{2})-H(\mathbf{Y}_{2}|\mathbf{X}_2)\nonumber\\
&=&H(\mathbf{X}_{1}|\mathbf{X}_{2})+H(W_{1},\mathbf{Y}_{2}|\mathbf{X}_{1},\mathbf{X}_{2})-H(\mathbf{X}_{1}|W_{1},\mathbf{Y}_{2},\mathbf{X}_{2})-H(\mathbf{Y}_{2}|\mathbf{X}_{2})\nonumber\\
&\geq&H(\mathbf{X}_{1})+H(\mathbf{Y}_{2}|\mathbf{X}_{1},\mathbf{X}_{2})-H(\mathbf{X}_{1}|W_{1},\mathbf{Y}_{2},\mathbf{X}_{2})-H(\mathbf{Y}_{2}|\mathbf{X}_{2}),
\end{eqnarray}
since $\mathbf{x}_1,\mathbf{x}_2$ are independent. This can then
be shown to be larger than
$n(I(X_1;Y|X_2)-I(X_1;Y_2|X_2)-\epsilon)$.

 If
$\min\{I(X_{2};Y),I(X_2;Y_2)\}=I(X_{2};Y_2)$, the last line
in~\eqref{eq:nf} changes to $R_e<
\big[I(X_{1};Y|X_{2})+\min\{I(X_{2};Y),I(X_{2};Y_{2}|X_{1})\}-I(X_{1},X_{2};Y_{2})\big]^+$,
then we can use a coding/decoding scheme similar to the one
developed above to show the achievability.

The claim is achieved.

\section{Proof of Theorem~\ref{thm:cf}}\label{ap:cf}
 The proof is a combination
of the coding scheme of Csisz$\acute{a}$r \emph{et.
al.}~\cite{Csiszar:TIT:78} and a revised CF scheme in the relay
channel~\cite{Cover:TIT:79}.

\begin{enumerate}
\item \textbf{Codebook generation:}

We first generate at random $2^{nR}$ i.i.d $n$-sequence
$\mathbf{x}_{1}$ at the source node each drawn according to
$p(\mathbf{x}_{1})=\prod\limits_{j=1}^{n}p(x_{1,j})$, index them
as $\mathbf{x}_{1}(k),k\in[1,2^{nR}]$, with
$R=I(X_1;\hat{Y}_1,Y|X_2)-\epsilon$.

Generate at random $2^{nR_{2}}$ i.i.d $n$-sequence
$\mathbf{x}_{2}$ each with probability
$p(\mathbf{x}_{2})=\prod\limits_{j=1}^{n}p(x_{2,j})$. Index these
as $\mathbf{x}_{2}(s),s\in[1,2^{nR_{2}}]$, where
$$R_2=\min\{I(X_2;Y),I(X_2;Y_2|X_1)\}-\epsilon.$$

For each $\mathbf{x}_{2}(s)$, generate at random $2^{n
(R_{2}-R_{0})}$ i.i.d $\mathbf{\hat{y}}_{1}$, each with
probability
$p(\hat{\mathbf{y}}_{1}|\mathbf{x}_{2}(s))=\prod\limits_{j=1}^{n}p(\hat{y}_{1,j}|x_{2,j}(s))$.
Label these
$\hat{\mathbf{y}}_{1}(z,s),z\in[1,2^{n\hat{R}}],s\in[1,2^{nR_{2}}]$,
where we set $\hat{R}=R_{2}-R_{0}$. Equally divide these
$2^{nR_{2}}$ $\mathbf{x}_{2}$ sequences into $2^{n\hat{R}}$ bins,
hence there are $2^{nR_{0}}$ $\mathbf{x}_{2}$ sequences at each
bin. Let $f$ be this mapping, that is $z=f(s)$.

Let
$R^{'}=\min\{I(X_{2};Y),I(X_{2};Y_{2}|X_{1})\}+I(X_{1};\hat{Y}_{1},Y|X_{2})-I(X_{1},X_{2};Y_{2})$.

Define $\mathcal{W}=\{1,\cdots,2^{nR^{'}}\}$,
$\mathcal{L}=\{1,\cdots,2^{n(R-R^{'})}\}$ and
$\mathcal{K}=\mathcal{W}\times\mathcal{L}=\{1,\cdots,2^{nR}\}$.
 \item \textbf{Encoding}

We exploit the block Markov coding scheme.

For a given rate pair $(R_{1},R_{e})$, where $R_1\leq R, R_e\leq
R_1$, we give the following coding strategy. Let the message to be
transmitted at block $i$ be $w_{1}(i)\in\mathcal{W}_{1}=[1,M]$,
where $M=2^{nR_{1}}$. We require $R_{1}\leq R$.

The stochastic encoder at the transmitter first forms the
following mappings.
\begin{itemize}

\item If $R_{1}>R^{'}$, we let
$\mathcal{W}_{1}=\mathcal{W}\times\mathcal{J}$, where
$\mathcal{J}=\{1,2^{n(R_{1}-R^{'})}\}$. We let $g_{1}$ be the
partition that partitions $\mathcal{L}$ into $|\mathcal{J}|$ equal
size subsets. The stochastic encoder at transmitter will choose a
mapping for each message
$w_{1}(i)=(w(i),j(i))\rightarrow(w(i),l(i))$, where $l(i)$ is
chosen randomly from the set $g_{1}^{-1}(j(i))\subset\mathcal{L}$
with uniform distribution.

\item If $R_{1}<R^{'}$, the stochastic encoder will choose a
mapping $w_{1}(i)\rightarrow (w_{1}(i),l(i))$, where $l(i)$ is
chosen uniformly from the set $\mathcal{L}$.
\end{itemize}

At first consider block $i$, where $i\neq 1,B$, which means it's
not the first or the last block. Assume that the message
$w_{1}(i)$ intended to send at block $i$ is associated with
$(w(i),l(i))$ by the stochastic encoder at the transmitter. We let
$k(i)=(w(i),l(i))$. Then the encoder at the source sends
$\mathbf{x}_{1}(k(i))$ at block $i$. At the end of block $i-1$, we
assume that
$(\mathbf{x}_{2}(s(i-1)),\hat{\mathbf{y}}_{1}(z(i-1),s(i-1)),\mathbf{y}_{1}(i-1))$
are jointly typical\footnote{See the decoding part, such $z(i-1)$
exists.}, then we choose $s(i)$ uniformly from bin $z(i-1)$, and
the relay sends $\mathbf{x}_{2}(s(i))$ at block $i$.

When $i=1$, the source sends $\mathbf{x}_{1}(k(1))$, the relay
sends $\mathbf{x}_{2}(1)$. When $i=B$, the source sends
$\mathbf{x}_{1}(1)$, the relay sends $\mathbf{x}_{2}(s(B))$.

\item \textbf{Decoding}

First consider the relay node. At the end of block $i$, the relay
already has $s(i)$\footnote{At the end of block $1$, relay knows
$s(i)=1$, this is the starting point.}, it then decides $z(i)$ by
choosing $z(i)$ such that
$(\mathbf{x}_{2}(s(i)),\hat{\mathbf{y}}_{1}(z(i),s(i)),\mathbf{y}_{1}(i))$
are jointly typical. There exists such $z(i)$, if
\begin{eqnarray}
\hat{R}\geq I(Y_{1};\hat{Y}_{1}|X_{2}),
\end{eqnarray}
and $n$ is sufficiently large. Choose $s(i+1)$ uniformly from bin
$z(i)$.

The destination does backward decoding. The decoding process
starts at the last block $B$, the destination decodes $s(B)$ by
choosing unique $\hat{s}(B)$ such that
$(\mathbf{x}_{2}(\hat{s}(B)),\mathbf{y}(B))$ are jointly typical.
We will have $\hat{s}(B)=s(B)$, if
\begin{eqnarray}\label{eq:x2bound}
R_{2}\leq I(X_{2};Y),
\end{eqnarray}
and $n$ is sufficiently large.

Next, the destination moves to the block $B-1$. Now it already has
$s(B)$, hence we also have $z(B-1)=f(s(B))$. It first declares
that $\hat{s}(B-1)$ is received, if $\hat{s}(B-1)$ is the unique
one such that $(\mathbf{x}_{2}(\hat{s}(B-1)),\mathbf{y}(B-1))$ are
jointly typical. If~\eqref{eq:x2bound} is satisfied,
$\hat{s}(B-1)=s(B-1)$ with high probability. After knowing
$\hat{s}(B-1)$, the destination gets an estimation of
$\hat{k}(B-1)$, by picking the unique $\hat{k}(B-1)$ such that
$(\mathbf{x}_{1}(\hat{k}(B-1)),\hat{\mathbf{y}}_{1}(z(B-1),\hat{s}(B-1)),\mathbf{y}(B-1),\mathbf{x}_{2}(\hat{s}(B-1)))$
are jointly typical. We will have $\hat{k}(B-1)=k(B-1)$ with high
probability, if
\begin{eqnarray}\label{eq:x1bound}
R\leq I(X_{1};\hat{Y}_{1},Y|X_{2}),
\end{eqnarray}
and $n$ is sufficiently large.

When the destination moves to block $i$, the destination has
$s(i+1)$ and hence $z(i)=f(s(i+1))$. It first declares that
$\hat{s}(i)$ is received, by choosing unique $\hat{s}(i)$ such
that $(\mathbf{x}_{2}(\hat{s}(i)),\mathbf{y}(i))$ are jointly
typical. If~\eqref{eq:x2bound} is satisfied, $\hat{s}(i)=s(i)$
with high probability. After knowing $\hat{s}(i)$, the destination
declares that $\hat{k}(i)$ is received, if $\hat{k}(i)$ is the
unique one such that
$(\mathbf{x}_{1}(\hat{k}(i)),\hat{\mathbf{y}}_{1}(z(i),\hat{s}(i)),$
$\mathbf{y}(i),\mathbf{x}_{2}(\hat{s}(i)))$ are jointly typical.
If~\eqref{eq:x1bound} is satisfied, $\hat{k}(i)=k(i)$ with high
probability when $n$ is sufficiently large.

Having $\hat{k}(i)$, the destination can get the estimation of the
message $w_{1}(i)$ by letting 1)
$\hat{w}_{1}(i)=(\hat{w}(i),\hat{j}(i))=(\hat{w}(i),g_{1}(\hat{l}(i)))$,
if $R_{1}>R-R^{'}$, 2) $\hat{w}_{1}(i)=\hat{w}(i)$, if
$R_{1}<R-R^{'}$. The probability that $\hat{w}_1(i)=w_1(i)$ goes
to one for sufficiently large $n$.

 \item
\textbf{Equivocation Computation}
\begin{eqnarray}
H(W_{1}|\mathbf{Y}_{2})&=&H(W_{1},\mathbf{Y}_{2})-H(\mathbf{Y}_{2})\nonumber\\
&=&H(W_{1},\mathbf{Y}_{2},\mathbf{X}_{1},\mathbf{X}_{2})-H(\mathbf{X}_{1},\mathbf{X}_{2}|W_{1},\mathbf{Y}_{2})-H(\mathbf{Y}_{2})\nonumber\\
&=&H(\mathbf{X}_{1},\mathbf{X}_{2})+H(W_{1},\mathbf{Y}_{2}|\mathbf{X}_{1},\mathbf{X}_{2})-H(\mathbf{X}_{1},\mathbf{X}_{2}|W_{1},\mathbf{Y}_{2})-H(\mathbf{Y}_{2})\nonumber\\
&\geq&H(\mathbf{X}_{1},\mathbf{X}_{2})+H(\mathbf{Y}_{2}|\mathbf{X}_{1},\mathbf{X}_{2})-H(\mathbf{X}_{1},\mathbf{X}_{2}|W_{1},\mathbf{Y}_{2})-H(\mathbf{Y}_{2}).
\end{eqnarray}

Following~\cite{Wyner:BSTJ:75}, we will have
$H(\mathbf{Y}_{2})-H(\mathbf{Y}_{2}|\mathbf{X}_{1},\mathbf{X}_{2})\leq
nI(X_{1},X_{2};Y_{2})+n\delta_n$, where $\delta_n\rightarrow 0$ as
$n\rightarrow \infty$.

Now let's calculate
$H(\mathbf{X}_{1},\mathbf{X}_{2}|W_{1},\mathbf{Y}_{2})$. Given
$W_{1}$, the eavesdropper can do joint decoding. It does backward
decoding. We pick up the story at block $i$, we suppose it already
decodes $s(i+1)$ and hence $z(i)=f(s(i+1))$. Given $W_{1}$, the
eavesdropper knows $w(i)$, hence it will decode $l(i)$ and $s(i)$
sent by the relay, by letting $l(i)=\hat{l}(i),s(i)=\hat{s}(i)$,
if $\hat{l}(i),\hat{s}(i)$ are the only ones such that
$(\mathbf{x}_{1}(w(i),\hat{l}(i)),\mathbf{x}_{2}(\hat{s}(i)),\mathbf{\hat{y}}_{1}(z(i),\hat{s}(i)),\mathbf{y}_{2}(i))$
are jointly typical. Then, if $R_{2}\leq I(X_{2};Y_{2}|X_{1})$
and~\eqref{eq:x1bound} is satisfied, we have
\begin{eqnarray}
\frac{1}{2}\log_{2}(|\mathcal{L}|)+R_{2}&=&R-R^{'}+R_{2}\nonumber\\
&=&R-\min\{I(X_{2};Y),I(X_{2};Y_{2}|X_{1})\}-I(X_{1};\hat{Y}_{1},Y|X_{2})\nonumber\\
&&+I(X_{1},X_{2};Y_{2})+\min\{I(X_{2};Y),I(X_{2};Y_{2}|X_{1})\}\nonumber\\
&\leq&I(X_{1},X_{2};Y_{2}).
\end{eqnarray}

Also, we have $$\frac{1}{2}\log_{2}(|\mathcal{L}|)<R\leq
I(X_{1};\hat{Y}_{1},Y_{2}|X_{2}).$$

Thus, we have
\begin{eqnarray}\text{Pr}\{(\mathbf{X}_{1}(w(i),\hat{l}(i)),\mathbf{X}_{2}(\hat{s}(i)))\neq
(\mathbf{X}_{1}(w(i),l(i)),\mathbf{X}_{2}(s(i)))\}\leq
\epsilon_{1}.\end{eqnarray}

Then based on Fano's inequality, we have
\begin{eqnarray}
\frac{1}{n}H(\mathbf{X}_{1},\mathbf{X}_{2},|W_{1}=w_{1},\mathbf{Y}_{2})\leq
\frac{1}{n}+\epsilon_{1} I(X_{1},X_{2};Y_{2}).
\end{eqnarray}
Hence, we have
\begin{eqnarray}
\frac{1}{n}H(\mathbf{X}_{1},\mathbf{X}_{2}|W_{1},\mathbf{Y}_{2})=\frac{1}{n}\sum\limits_{w_{1}\in\mathcal{W}_{1}}p(W_{1}=w_{1})H(\mathbf{X}_{1},\mathbf{X}_{2}|W_{1}=w_{1},\mathbf{Y}_{2})\leq
\epsilon_{2},
\end{eqnarray}
 when $n$ is sufficiently large.

We know
$H(\mathbf{X}_{1},\mathbf{X}_{2})=H(\mathbf{X}_{1})+H(\mathbf{X}_{2}|\mathbf{X}_{1})\geq
n(R+R_{0})$.

If $R_1>R^{'}$, we have $H(\mathbf{X}_{1})=nR$, then we get
$$nR_{e}=H(W_{1}|\mathbf{Y}_{2})\geq
n(R_{0}+I(X_{1};\hat{Y}_{1},Y|X_{2})-I(X_{1},X_{2};Y_{2})-\epsilon_{4}).$$
If $R_1<R^{'}$, we have $H(\mathbf{X}_{1})=n(R_1+R-R^{'})$, hence
$$nR_{e}=H(W_{1}|\mathbf{Y}_{2})\geq
nR_{1}+n(R_{0}-\min\{I(X_2;Y),I(X_2;Y_2|X_1))\}-\epsilon_{4}).$$

\end{enumerate}

The claim is proved.

\section{Proof of Theorem~\ref{thm:nfdeaf}}\label{ap:nfdeaf}

The proof follows closely with that of Theorem~\ref{thm:nf}. We
first consider the case $I(X_{1};Y|X_{2})<I(X_{1};Y_{2}|X_{2})$,
\emph{i.e.}, the channel between the source and the eavesdropper
is better than the channel between the source and the destination.
In this case, we only need to consider the case $\min\{I(X_2;Y),
I(X_2;Y_2)\}=I(X_2;Y_2)$, otherwise, the perfect secrecy rate will
be zero. Thus in this case,  $R_{s1}=
\big[I(X_{1};Y|X_{2})+\min\{I(X_{2};Y),I(X_{2};Y_{2}|X_{1})\}-I(X_{1},X_{2};Y_{2})\big]^+$.
\begin{enumerate}
\item \textbf{Codebook generation:}

For a given distribution $p(x_{1})p(x_{2})$, we generate at random
$2^{nR_{2}}$ i.i.d $n$-sequence at the relay node each drawn
according to $p(\mathbf{x}_{2})=\prod_{i=1}^{n}p(x_{2,i})$, index
them as $\mathbf{x}_{2}(a),a\in[1,2^{nR_{2}}]$. Here we set
$R_2=\min\{I(X_2;Y),I(X_2;Y_2|X_1)\}-\epsilon$. We also generate
random $2^{nR}$ i.i.d $n$-sequence at the source each drawn
according to $p(\mathbf{x}_{1})=\prod_{i=1}^{n}p(x_{1,i})$, index
them as $\mathbf{x}_{1}(k),k\in[1,2^{nR}]$ with
$R=I(X_1;Y|X_2)-\epsilon$. Let
$$R_{min}=\min\{R_{s1},R_{s2}\},R_{max}=\max\{R_{s1},R_{s2}\},$$
where
$R_{s1}=\min\{I(X_{2};Y),I(X_{2};Y_{2}|X_{1})\}+I(X_{1};Y|X_{2})-I(X_{1},X_{2};Y_{2}),$
$R_{s2}=I(X_{1};Y|X_2)-I(X_{1};Y_1|X_2)$. We now define
$\mathcal{W} = \{1,\cdots,2^{nR_{min}}\},$
$\mathcal{L}_{1}=\{1,\cdots,2^{n(R_{max}-R_{min})}\}$,
$\mathcal{L}_{2}=\{1,\cdots,2^{n(R-R_{max})}\}$ and
$\mathcal{L}=\mathcal{L}_1\times\mathcal{L}_2,$
$\mathcal{K}=\mathcal{W}\times\mathcal{L}$.

\item \textbf{Encoding}

Here, we consider perfect secrecy rate. For a given rate
$R_{1}\leq R_{min}$, we give the following coding strategy to show
that for any given $\epsilon\geq 0$, the equivocation rate at the
eavesdropper and the relay node can be made to be larger or equal
$R_{1}-\epsilon$ .

Let the message to be transmitted at block $i$ be
$w_{1}(i)\in\mathcal{W}_{1}=[1,M]$, where $M=2^{nR_{1}}$. The
stochastic encoder will choose a mapping $w_{1}(i)\rightarrow
(w_{1}(i),l_1(i),l_2(i))$, where $l_1(i),l_2(i)$ are chosen
uniformly from the set $\mathcal{L}_1,\mathcal{L}_2$ respectively.
We write $l(i)=(l_1(i),l_2(i))$.

Suppose the message $w_{1}(i)$ intended to send at block $i$ is
associated with $(w(i),l(i))$
 by the stochastic encoder at the transmitter. The
encoder then sends $\mathbf{x}_{1}((w(i),l(i)))$. The relay
uniformly picks a code $\mathbf{x}_{2}(a)$ from
$a\in[1,\cdots,2^{nR_{2}}]$, and sends $\mathbf{x}_{2}(a)$.

\item \textbf{Decoding}

At the end of block $i$, the destination declares that
$\hat{a}(i)$ is received, if $\hat{a}(i)$ is the only one such
that $(\mathbf{x}_{2}(\hat{a}(i)),\mathbf{y})$ are jointly
typical. If there does not exist or there exist more than one such
sequences, the destination declares an error. Since
$R_2=\min\{I(X_{2};Y),I(X_{2};Y_2|X_{1})\}-\epsilon\leq
I(X_{2};Y)-\epsilon$, then based on AEP, we know that the error
probability will be less than any given positive number
$\epsilon$, when the codeword length $n$ is long enough.

The destination then declares that $\hat{k}$ is received, if
$\hat{k}$ is the only one such that
$(\mathbf{x}_{1}(\hat{k}),\mathbf{x}_{2}(\hat{a}),\mathbf{y})$ are
jointly typical, otherwise declares an error. Since
$R=I(X_{1};Y|X_{2})-\epsilon$, then based on AEP, we know that we
will have error probability goes to zero, when $n$ is sufficiently
large.

Having $\hat{k}(i)$, the destination can get the estimation of the
message $w_{1}(i)$ by letting $\hat{w}_{1}(i)=\hat{w}(i)$. The
probability that $\hat{w}_1(i)=w_1(i)$ goes to one for
sufficiently large $n$.

 \item
\textbf{Equivocation Computation}

We first calculate the equivocation rate of the eavesdropper when
$R_{s1}\leq R_{s2}$.
\begin{eqnarray}
H(W_{1}|\mathbf{Y}_{2})&=&H(W_{1},\mathbf{Y}_{2})-H(\mathbf{Y}_{2})\nonumber\\
&=&H(W_{1},\mathbf{Y}_{2},\mathbf{X}_{1},\mathbf{X}_{2})-H(\mathbf{X}_{1},\mathbf{X}_{2}|W_{1},\mathbf{Y}_{2})-H(\mathbf{Y}_{2})\nonumber\\
&=&H(\mathbf{X}_{1},\mathbf{X}_{2})+H(W_{1},\mathbf{Y}_{2}|\mathbf{X}_{1},\mathbf{X}_{2})-H(\mathbf{X}_{1},\mathbf{X}_{2}|W_{1},\mathbf{Y}_{2})-H(\mathbf{Y}_{2})\nonumber\\
&\geq&H(\mathbf{X}_{1},\mathbf{X}_{2})+H(\mathbf{Y}_{2}|\mathbf{X}_{1},\mathbf{X}_{2})-H(\mathbf{X}_{1},\mathbf{X}_{2}|W_{1},\mathbf{Y}_{2})-H(\mathbf{Y}_{2}).
\end{eqnarray}

Now let's calculate
$H(\mathbf{X}_{1},\mathbf{X}_{2}|W_{1},\mathbf{Y}_{2})$. Given
$W_{1}$, the eavesdropper can do joint decoding. At any block $i$,
given $W_{1}$, the eavesdropper knows $w(i)$, hence it will decode
$l(i)=(l_1(i),l_2(i))$ and $a(i)$ sent by the relay, by letting
$l(i)=\hat{l}(i),a(i)=\hat{a}(i)$, if $\hat{l}(i),\hat{a}(i)$ are
the only one pair such that
$(\mathbf{x}_{1}(w(i),\hat{l}(i)),\mathbf{x}_{2}(\hat{a}(i)),\mathbf{y})$
are jointly typical. Since
$R_2=\min\{I(X_{2};Y),I(X_{2};Y_2|X_{1})\}-\epsilon\leq
I(X_{2};Y_2|X_{1})-\epsilon$, we
\begin{eqnarray}
\frac{1}{2}\log_{2}(|\mathcal{L}|)+R_{2}&=&R+I(X_{1},X_{2};Y_{2})-\min\{I(X_{2};Y),I(X_{2};Y_{2}|X_{1})\}\nonumber
\\&&-I(X_{1};Y|X_{2})+\min\{I(X_{2};Y),I(X_{2};Y_{2}|X_{1})\}-\epsilon\nonumber
\\&\leq&I(X_{1},X_{2};Y_{2})-\epsilon.
\end{eqnarray}
Also, we have $\frac{1}{2}\log_{2}(|\mathcal{L}|)<R\leq
I(X_{1};Y_{2}|X_{2}).$

So
$\text{Pr}\{(\mathbf{X}_{1}(w(i),\hat{l}(i)),\mathbf{X}_{2}(\hat{a}(i)))\neq
(\mathbf{X}_{1}(w(i),l(i)),\mathbf{X}_{2}(a(i)))\}\leq
\epsilon_{1}.$

Then based on Fano's inequality, we have
\begin{eqnarray}
\frac{1}{n}H(\mathbf{X}_{1},\mathbf{X}_{2}|W_{1}=w_{1},\mathbf{Y}_{2})\leq
\frac{1}{n}+\epsilon_{1} I(X_{1},X_{2};Y_{2}).
\end{eqnarray}
Hence, we have
\begin{eqnarray}
\frac{1}{n}H(\mathbf{X}_{1},\mathbf{X}_{2}|W_{1},\mathbf{Y}_{2})=\frac{1}{n}\sum\limits_{w_{1}\in\mathcal{W}_{1}}p(W_{1}=w_{1})H(\mathbf{X}_{1},\mathbf{X}_{2}|W_{1}=w_{1},\mathbf{Y}_{2})\leq
\epsilon_{2},
\end{eqnarray}
 when $n$ is sufficiently large.

Now,
$H(\mathbf{Y}_{2})-H(\mathbf{Y}_{2}|\mathbf{X}_{1},\mathbf{X}_{2})\leq
nI(X_{1},X_{2};Y_{2})+n\delta_n,$ where $\delta_n\to 0,$ as $n\to
\infty$. Also we have
$H(\mathbf{X}_{1},\mathbf{X}_{2})=H(\mathbf{X}_{1})+H(\mathbf{X}_{2})$
since $\mathbf{x}_{1}$ and $\mathbf{x}_{2}$ are independent. Now
$H(\mathbf{X}_1)=R_{1}+I(X_1,X_2;Y_2)-\min\{I(X_2;Y),I(X_2;Y_2|X_1)\},$
hence
$H(\mathbf{X}_{1},\mathbf{X}_{2})=R_1+I(X_1,X_2;Y_2)-\epsilon$.

We get $nR_{e}=H(W_{1}|\mathbf{Y}_{2})\geq nR_1-n\epsilon_4$.
\end{enumerate}

Now we calculate the equivocation rate at the relay node.
\begin{eqnarray}
H(W_{1}|\mathbf{Y}_1,\mathbf{X}_2)&\geq&H(W_{1}|\mathbf{Y}_1,\mathbf{X}_2,L_1)\no\\
&=&H(W_{1},\mathbf{Y}_1,L_1|\mathbf{X}_2)-H(\mathbf{Y}_1,L_1|\mathbf{X}_2)\no\\
&=&H(W_{1},L_{1},\mathbf{Y}_1,\mathbf{X}_1|\mathbf{X}_2)-H(\mathbf{X}_1|W_{1},L_{1},\mathbf{Y}_1,\mathbf{X}_2)-H(\mathbf{Y}_1,L_1|\mathbf{X}_2)\no\\
&=&H(\mathbf{X}_1|\mathbf{X}_2)+H(W_{1},L_{1},\mathbf{Y}_1|\mathbf{X}_1,\mathbf{X}_2)-H(\mathbf{X}_1|W_{1},L_{1},\mathbf{Y}_1,\mathbf{X}_2)-H(\mathbf{Y}_1,L_1|\mathbf{X}_2)\no\\
&\overset{(a)}{\geq}&H(\mathbf{X}_1)+H(\mathbf{Y}_1|\mathbf{X}_1,\mathbf{X}_2)-H(\mathbf{X}_1|W_{1},L_{1},\mathbf{Y}_1,\mathbf{X}_2)-H(L_1)-H(\mathbf{Y}_1|\mathbf{X}_2),\no
\end{eqnarray}
where the first term of (a) comes from the fact that
$\mathbf{x}_1,\mathbf{x}_2$ are independent, and the fourth term
comes from the fact that $l_1,\mathbf{x}_2$ are independent.

Now, $H(L_1)=n(R_{max}-R_{min})$,
$H(\mathbf{Y}_1|\mathbf{X}_1,\mathbf{X}_2)-H(\mathbf{Y}_1|\mathbf{X}_2)\leq
nI(X_1;Y_1|X_2)+n\delta_n$. Given $w_{1},l_1,\mathbf{x}_{2}$, the
relay can just choose the $\mathbf{x}_1$ in the bin $(w_{1},l_1)$
which is jointly typical with $\mathbf{x}_2,\mathbf{y}_1$. Since
$\frac{1}{n}\log_2(|\mathcal{L}_2|)\leq I(X_1;Y_1|X_2)$, we have
$\text{Pr}\{\hat{\mathbf{X}}_1\neq \mathbf{X}_1\}\leq \epsilon_2$.

Then based on Fano's inequality, we have
\begin{eqnarray}
\frac{1}{n}H(\mathbf{X}_{1}|W_{1}=w_{1},L_{1}=l_1,
\mathbf{Y}_{1},\mathbf{X}_2=\mathbf{x}_{2})\leq
\frac{1}{n}+\epsilon_{1} I(X_{1};Y_{1}|X_2),
\end{eqnarray}
Hence, we have
\begin{eqnarray}
\frac{1}{n}H(\mathbf{X}_{1}|W_{1},L_1,\mathbf{Y}_{1},\mathbf{X}_2)&=&\frac{1}{n}\sum\limits_{w_{1},l_{1},\mathbf{x}_2}p(W_{1}=w_{1},L_{1}=l_1,\mathbf{x}_2)H(\mathbf{X}_{1},\mathbf{X}_{2}|W_{1}=w_{1},L_{1}=l_{i},\mathbf{x}_2,\mathbf{Y}_{1})\no\\&\leq&
\epsilon_{2},
\end{eqnarray}
 when $n$ is sufficiently large.

Also, based on the encoding part, we have
$H(\mathbf{X}_1)=n(R_{1}+I(X_1;Y|X_2)-R_{min})$.

Combining these, we get
\begin{eqnarray}
H(W_{1}|\mathbf{Y}_1,\mathbf{X}_2)&\geq&n(R_{1}+I(X_1;Y|X_2)-R_{min}-(R_{max}-R_{min})-I(X_1;Y_1|X_2)-\delta_n)\no\\
&=&n(R_1-\delta_n).
\end{eqnarray}

The equivocation rate of the relay and the eavesdropper when
$R_{1s}\geq R_{2s}$ can be calculated similarly, with the only
difference that we bound the equivocation rate of the eavesdropper
by giving it $L_1$.
 This case is proved.

Now, consider the case $I(X_{1};Y|X_{2})>I(X_{1};Y_{2}|X_2)$. If
$\min\{I(X_{2};Y),I(X_2;Y_2)\}=I(X_{2};Y)$, then we have
$\min\{I(X_{2};Y),I(X_2;Y_2|X_1)\}=I(X_{2};Y)$, because
$I(X_2;Y_2|X_1)>I(X_2;Y_2)$ since $X_1,X_2$ are independent. Under
this case, we only need to prove the case
$R_{s1}=\big[I(X_1;Y|X_2)-I(X_1;Y_2|X_2)\big]^+$, which can be
achieved by using a scheme similar to the one developed in
proving~\eqref{eq:ocase}. If $\min\{I(X_{2};Y),$
$I(X_2;Y_2)\}=I(X_{2};Y_2)$, and we only need to consider $R_{s1}=
\big[I(X_{1};Y|X_{2})+\min\{I(X_{2};Y),$
$I(X_{2};Y_{2}|X_{1})\}-I(X_{1},X_{2};Y_{2})\big]^+$, then we can
use a coding/decoding scheme similar to the one developed above to
show the achievability.

The claim is achieved.

 \vspace{-5mm}


\end{document}